\documentclass[showpacs,amsmath,pre]{revtex4}
\usepackage[pdftex]{graphicx}
\pdfoutput=1
\usepackage{amsmath}
\usepackage{dcolumn}

\usepackage{epsfig}
\usepackage{graphics}
\usepackage{latexsym}
\usepackage{amsfonts}
\usepackage{amssymb}

\begin{document}

\title{A statistical mechanical description of  metastable states and hysteresis in the 3D  soft-spin random-field model at $T=0$}
\author{M.L. Rosinberg and G. Tarjus}
\affiliation{Laboratoire de Physique Th\'eorique de la Mati\`ere Condens\'ee, Universit\'e Pierre et Marie Curie\\ 4 place Jussieu, 75252 Paris Cedex 05, France}

\begin{abstract}
We present a formalism for computing the complexity of metastable states and the zero-temperature magnetic hysteresis loop in the soft-spin random-field model in finite dimensions. The complexity is obtained as the Legendre transform of the free-energy associated to a certain action in replica space and the hysteresis loop above the critical disorder is defined as the curve in the field-magnetization plane where the complexity vanishes; the nonequilibrium magnetization is therefore obtained without having to follow  the dynamical evolution. We use approximations borrowed from condensed-matter theory and based on assumptions on the structure of the direct correlation functions (or proper vertices), such as a local approximation for the self-energies, to calculate the hysteresis loop in three dimensions, the correlation functions along the loop, and the second moment of the avalanche-size distribution.
\end{abstract}

\pacs{75.10.Nr, 75.60.Ej, 64.60.av}

\maketitle

\def\be{\begin{equation}}
\def\ee{\end{equation}}
\def\bea{\begin{align}}
\def\eea{\end{align}}

\section{Introduction}
The random-field Ising model (RFIM) at zero temperature is a  prototype of many disordered systems which exhibit hysteretic and jerky behavior when an external parameter ({\it e.g.} magnetic field, pressure, strain) is slowly changed\cite{SDP2006}. This behavior is related to the existence of a corrugated (free) energy  landscape with many local minima (or metastable states) separated by barriers much larger than $k_B T$, therefore preventing relaxation towards equilibrium on experimental time scales. As a consequence, the response to an external  driving field is a series of discontinuous jumps (called avalanches) from one metastable state to another, the number and size of these jumps varying with the amount of disorder. In  the 3-dimensional RFIM with Gaussian random fields two regimes of avalanches are observed\cite{S1993,DS1996}:  at large disorder, there are many microscopic jumps resulting in a smooth magnetization curve macroscopically; at low disorder, there is a system-spanning avalanche resulting in a  macroscopic jump in the magnetization curve.  These two regimes are separated by a critical point at which avalanches of all sizes are observed.  As shown recently, this type of nonequilibrium disorder-induced phase transition may underly the hysteretic behavior of $^4$He adsorbed in high porosity silica aerogels\cite{DKRT2005, BLCGDPW2008}.

Even though the $T=0$ RFIM and similar models have been extensively studied, there is currently no analytical tool to compute the saturation hysteresis loop in finite dimensions (even approximately) and no theory to describe the statistical properties of the metastable states, for instance their configurational entropy  (also called `complexity') as a function of magnetization. Furthermore, the behavior of the  correlation functions along the hysteresis loop is unknown, although this is an issue of experimental interest\cite{DKRT2006}.
Of course, describing analytically such a nonequilibrium situation where the present state of the system depends on its past history is  a difficult problem and at first sight there seems to be no other choice than to follow the dynamical evolution for given initial conditions ({\it e.g.} one of the two saturated states corresponding to infinitely positive or negative  magnetic field). In this way, one can treat exactly  `mean-field' systems, {\it i.e.} fully connected lattices\cite{S1993,DS1996} or lattices with a locally tree-like structure such as the Bethe lattice\cite{DSS1997,SSD2000}, and obtain analytical expressions for the saturation hysteresis loop or the avalanche size distribution. To go further and build a field-theoretical description of these phenomena it then seems necessary to employ the Martin-Siggia-Rose formalism\cite{DS1996}. In ferromagnetic systems, however, an alternative route is possible thanks to a remarkable property of the saturation hysteresis loop in the large-disorder regime: the loop is the convex hull of the metastable states in the two-dimensional field-magnetization plane\cite{DRT2005,PRT2008,RTP2009,RM2009}. In other words, the complexity is zero outside the loop and positive inside\cite{note1}. Moreover, it exactly vanishes along the loop since there is a single `extremal' metastable state at a given field (at least for a continuous distribution of the random field). Determining the hysteresis loop is then tantamount to counting the number of metastable states at fixed field as a function of their magnetization and finding the magnetization at which the complexity vanishes. The main difficulty is that one must count the {\it typical} ({\it i.e.} most likely) number of metastable states and compute the associated  {\it quenched} complexity, which of course is a nontrivial task requiring the use of replicas. Nevertheless, at least in principle, one can thereby study hysteresis and avalanche statistics without referring to the dynamical evolution. 

In the present paper we test this approach by studying the  soft-spin version of the RFIM introduced in Ref.\cite{DS1996}, where the spins take on continuous values between $-\infty$ and $+\infty$ and are confined by a bistable  potential with a linear cusp. This model has the advantage over the standard RFIM to show hysteresis for all values of the disorder in mean-field theory and is anyhow expected to behave as the hard-spin model on long length scales. Our basic strategy is to express the complexity as the Legendre transform of the `free-energy' associated to a certain action in replica space and then to use approximations borrowed from condensed-matter theory to calculate the corresponding correlation (or Green's) functions and thermodynamic properties. Our focus on the correlation functions is also motivated by analytical and numerical calculations on the Bethe lattice which suggest that their spatial structure along the hysteresis loop is simpler than at equilibrium\cite{IR2010}.

The paper is organized as follows. In section II, we define the model and the observables that we want to compute. In section III, we introduce the general formalism in the context of a replica method, focusing on the various correlation functions. In section IV, we first consider the random-phase approximation (RPA) which is equivalent to mean-field theory and becomes exact in infinite dimension. In section V, we go beyond the RPA by introducing a local self-energy approximation (LSEA) and we compare the predictions for the magnetization, the correlation functions, and the second-moment of the avalanche-size distribution along the hysteresis loop to simulation data. Concluding remarks and directions for future work are provided in Section VI. Additional details on the analytical calculations are provided in the appendix.

\section{Model and observables}

We consider a collection of $N$ soft spins placed on the sites of a $d$-dimensional hypercubic lattice and interacting via the Hamitonian 
\begin{equation}
\label{Eq1}
 {\cal H} = -J\sum_{<i,j>}s_is_j-\sum_i(H+h_i)s_i+\sum_i V(s_i)
\end{equation}
where $J>0$ is a ferromagnetic interaction that couples nearest-neighbor spins, $H$ is an external uniform field, and $\{h_i\}$ is a set of quenched random fields drawn independently from a Gaussian probability distribution  $p(h)$ with zero mean and variance $\Delta$. $V(s)$ is a double-well potential for which we choose the same cuspy form as in Ref.\cite{DS1996,RM2009} 
\begin{equation}
\label{Eq2}
V(s) = \frac{k}{2}[s-\mbox{sign}(s)]^2 
\end{equation}
so that all solutions of $\partial {\cal H}/\partial s_i=0$, {\it i.e.} all stationary points of the Hamiltonian, are local minima. This greatly facilitates the present study as there is no need to explicitly  discard local maxima and saddle-points through the consideration of the Hessian of the energy function. These local minima are the so-called  `metastable' states in which each spin satisfies 
\begin{equation}
\label{Eq3}
s_i-\mbox{sign}(s_i) = \frac{H+J\sum_{j/i} s_j+h_i}{k} 
\end{equation}
where $j$ is a neighbor of $i$ on the lattice.  
\begin{figure}[hbt]
\begin{center}
\includegraphics[width=8cm]{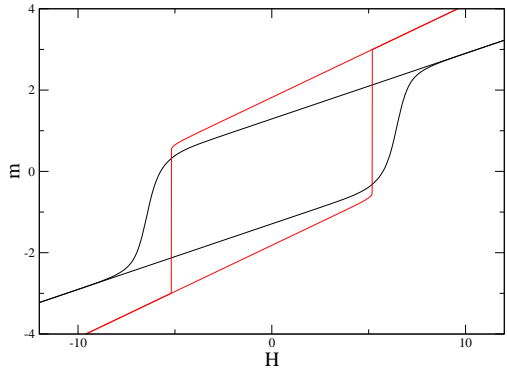}
\caption{\label{Fig1} Magnetization curves for the nonequilibrium zero-temperature random-field soft-spin model on a cubic lattice for $k=8$, $\Delta=4$, $J=0.3$ (black) and $J=0.6$ (red). The simulation data correspond to a single disorder realization of linear size $L=100$ using an increment in the external field $\delta H=0.01$.}
\end{center}
\end{figure}

At zero-temperature, one can define a local relaxation dynamics in which each spin is forced to satisfy Eq. \ref{Eq3} as the external field  is changed\cite{DS1996}. When adiabatically varying $H$ from $-\infty$ to $+\infty$ and backwards, the model exhibits hysteresis and the magnetization typically behaves as shown in Fig. 1. (The figure displays the results of a single simulation on a cubic lattice with $k=8$ and $\Delta=4$\cite{note5}.) These values are arbitrarily chosen and will stay fixed in the rest of this work. The shape of the loop then changes with the coupling $J$: one can see that $J=0.3$ and $J=0.6$ correspond to the large- and small-disorder regimes respectively, with a macroscopic  jump in the latter case.

For a given realization of the disorder, {\it i.e.} a set of random fields $\mathbf{h}=\{h_i\}$, and a given value of the external uniform field $H$, each metastable state is  characterized at a macroscopic level by  its magnetization,  its energy, etc. In the present study,  we only focus on the magnetization since this is sufficient to unambiguously determine the hysteresis loop. Our goal is then  to compute properties averaged over all metastable states with a given magnetization $m$ per site at a given external magnetic field $H$ (with a flat measure). A central quantity is the quenched complexity $\Sigma_Q(m,H)$, which is defined as
\begin{align}
\label{Eq4}
 \Sigma_Q(m,H)=\lim_{N \rightarrow \infty}\frac{1}{N} \overline {\ln {\cal N}(m,H;\mathbf{h})}
\end{align}
where  ${\cal N}(m,H;\mathbf{h})$ is the  number of metastable states with magnetization $m$ at the field $H$ and the overbar denotes an average over the random-field distribution. We are also interested in the following two-point correlation (Green's) functions:

(i) the spin-spin correlation functions,
\begin{align}
\label{Eq_spin_spin}
G_{ss,ij}(m,H)=G_{ij}(m,H)= G_{c,ij}(m,H) + G_{d,ij}(m,H),
\end{align}
with the (disorder) connected and disconnected components defined as
\begin{align}
\label{Eq_spin_spin_conn}
G_{c,ij}(m,H)=\overline{<s_i s_j>-<s_i><s_j>}
\end{align}
and
\begin{align}
\label{Eq_spin_spin_disc}
G_{d,ij}(m,H)=\overline{<s_i> < s_j>}-\overline{<s_i>} \ \ \overline{<s_j>} \ ,
\end{align}
the brackets denoting an averaged over all metastable states with magnetization $m$ at the field $H$,

(ii) the correlation  function involving the spin variable and the random field,
\begin{align}
\label{Eq_spin_field}
G_{sh,ij}(m,H)=\overline{<s_i> h_j} \ ,
\end{align}

(iii) the spin-spin correlation function for two copies of the same disordered system coupled to different external fields and with different magnetizations,
\begin{align}
\label{Eq_spin_spin_2copies}
G_{d,ij}(m^a,H^a;m^b,H^b)=\overline{<s_i>_a < s_j>_b}-\overline{<s_i>_a} \ \ \overline{<s_j>_b} \ ,
\end{align}
where the subscrit $a$ indicates that $\{m^a,H^a\}$ are fixed and similarly for the subscript $b$.

Along the hysteresis loop, the complexity is zero (at least in the large disorder regime where the loop is continuous)\cite{DRT2005,PRT2008,RTP2009,RM2009} and the connected spin-spin correlation is identically zero, as the fluctuations inside a metastable state vanish at zero temperature. On the other hand, the disconnected spin-spin correlation function and the spin-random-field one should be nontrivial.

Finally, we are also interested in the distribution of avalanche sizes along the loop (say, along the ascending branch). For a given disorder realization, the magnetization curve $m(H;\mathbf{h}) =(1/N)\sum_i s_i$  consists of  smooth parts resulting from the harmonic response to the magnetic field and a series of jumps of size $S_{\alpha}(\mathbf{h})$  occurring at the fields $H_{\alpha}(\mathbf{h})$  (these jumps are not visible in Fig. 1 due to the scale of the figure) \cite{note2}. The magnetization can thus be decomposed as
\begin{align}
\label{magcusp}
m(H;\mathbf{h})=m^{smooth}(H;\mathbf{h})+\sum_{\alpha}S_{\alpha}\theta(H-H_{\alpha})
\end{align}
where $\theta(x)$ is the Heaviside step function. From this, one defines the jump (avalanche) density 
\begin{align}
\rho(S,H)=\overline{\sum_{\alpha}\delta(S-S_{\alpha})\delta(H-H_{\alpha})},
\end{align}
so that $\rho(S,H)dSdH$ is the number of avalanches of size between $S$ and $S+dS$ when the field is increased from $H$ to $H+dH$ (note that $S$ is measured per site).  In present work we will only compute the `unnormalized' second moment $\int S^2\rho(S,H) dS=\overline{\sum_{\alpha}S_{\alpha}^2\delta(H-H_{\alpha})}$ (it is `unnormalized' because $\rho(S,H)$ as such is not a probability density and $\int \rho(S,H) dS$ is the total number of avalanches between $H$ and $H+dH$). 

\section{Formalism}

In order to control the local magnetization, we introduce an additional source $\mathbf{\hat{H}}=\{\hat{H}_i\}$ that is linearly coupled to the spins ${\bf s}=\{s_i\}$ and we consider the following (disorder-dependent) `partition function' in an external magnetic  field which is momentarily taken as nonuniform, $\mathbf{H}=\{H_i\}$:
\begin{align}
\label{Eq4}
{\cal Z}[\mathbf{H}, \mathbf{\hat{H}};\mathbf{h}]&= \int {\cal D}s\ e^{\mathbf{\hat{H}}.\mathbf{s}}\prod_i\delta\Big(\frac{\partial {\cal H}}{\partial s_i}\Big)\nonumber\\
 &=\int {\cal D}s\ e^{\mathbf{\hat{H}}.\mathbf{s}}\prod_i\delta (V'(s_i)-J\sum_{j/i}s_j-H_i-h_i) 
\end{align}
where the symbol ${\cal D}s$ refers to the integration over all the spin variables, ${\cal D}s=ds_1...ds_N$, and no Jacobian is needed (it would merely introduce a constant multiplicative factor). This partition function can be put into a more standard form by replacing the Dirac delta function by its Fourier representation, 
\begin{align}
\label{Eq4a}
{\cal Z}[\mathbf{H}, \mathbf{\hat{H}};\mathbf{h}]= \int  {\cal D}s  {\cal D}\hat{s}\ e^{-S[\mathbf{s},\hat{\mathbf{s}},\mathbf{h}]+\mathbf{\hat{H}}.\mathbf{s}+\mathbf{H}.\hat{\mathbf{s}}}
\end{align}
where the action $S$ is defined by
\begin{equation}
\label{Eq5}
S[\mathbf{s},\hat{\mathbf{s}},\mathbf{h}]= \sum_i \hat{s}_i [V'(s_i) -J\sum_{j/i}s_j-h_i]  
\end{equation}
and $\hat{\mathbf{s}}=\{\hat{s_i}\}$ are auxiliary (imaginary) variables; for conciseness, the factor $1/(2i\pi)$ associated to the integration of $\hat{s}_i$ along the imaginary axis is adsorbed into $d\hat{s}_i$. 

This defines a `free energy'  $W[\mathbf{H},\mathbf{\hat{H}};\mathbf{h}]=\ln {\cal Z}[\mathbf{H},\mathbf{\hat{H}};\mathbf{h}]$ which is a random object whose cumulants give access to full information about the system, including the complexity and the correlation functions. The complexity is obtained from the first cumulant,
\begin{align}
W_1[\mathbf{H}, \mathbf{\hat{H}}]=\overline{W[\mathbf{H}, \mathbf{\hat{H}};\mathbf{h}]},
\end{align} 
via a Legendre transform, where
\begin{align}
m_i[\mathbf{H}, \mathbf{\hat{H}}]= \frac{\partial W_1[\mathbf{H}, \mathbf{\hat{H}}]}{\partial {\hat H}_i} ,
\end{align}
and which for uniform sources takes the form
 \begin{align}
\label{eq_legendre_uniform}
\Sigma_Q(m,H) = \frac{1}{N} W_1(H,\hat{H}) - m(H,\hat{H})\hat{H}.
\end{align}
(Here and below we use square brackets $[...]$ when the arguments are locally varying and parenthesis $(...)$ when they are uniform.) As discussed in Refs.\cite{DRT2005,PRT2008,RTP2009,RM2009}, the hysteresis loop in the large-disorder regime identifies with the curve $\Sigma_Q(m,H) =0$ in the limit $\hat{H} \rightarrow \pm \infty$ whereas the typical properties of the metastable states are obtained for $\hat{H}=0$ which corresponds to the maximum of the complexity. 

The information about the distribution of avalanche sizes is contained in the higher-order cumulants $W_2[\mathbf{H}^1, \mathbf{\hat{H}}^1;\mathbf{H}^2, \mathbf{\hat{H}}^2],$ $W_3[\mathbf{H}^1, \mathbf{\hat{H}}^1;\mathbf{H}^2, \mathbf{\hat{H}}^2;\mathbf{H}^3, \mathbf{\hat{H}}^3]$,..., where 
\begin{align}
W_2[\mathbf{H}^1, \mathbf{\hat{H}}^1;\mathbf{H}^2, \mathbf{\hat{H}}^2]=\overline{W[\mathbf{H}^1, \mathbf{\hat{H}}^1;\mathbf{h}] W[\mathbf{H}^2, \mathbf{\hat{H}}^2;\mathbf{h}]}-\overline{W[\mathbf{H}^1, \mathbf{\hat{H}}^1;\mathbf{h}]} \ \overline{W[\mathbf{H}^2, \mathbf{\hat{H}}^2;\mathbf{h}]}
\end{align}
etc...
 
The correlation (Green's) functions are obtained by derivation of the cumulants with respect to the sources. For instance, the physical two-point spin-spin correlation functions introduced above are given by
\begin{align}
G_{c,ij}= \frac{\partial W_1[\mathbf{H},\mathbf{\hat{H}}]}{\partial \hat{H}_i\partial \hat{H}_j},
\end{align}
\begin{align}
\label{EqGd}
G_{d,ij}=\frac{\partial W_2[\mathbf{H}^1,\mathbf{\hat{H}}^1,\mathbf{H}^2,\mathbf{\hat{H}}^2]}{\partial \hat{H}_{i}^1\partial \hat{H}_{j}^2}\vert_{\mathbf{H}^1=\mathbf{H}^2=\mathbf{H},\mathbf{\hat{H}}^1=\mathbf{\hat{H}}^2=\mathbf{\hat{H}}}\  ,
\end{align}
where both right-hand sides are evaluated for uniform sources and $\hat{H}$ is considered as a function of $m$ and $H$ through the Legendre transform in Eq.~(\ref{eq_legendre_uniform}). The spin-random-field correlation function requires a little more thought. By using the property of Gaussian distributions,
\begin{align}
\label{Eq10a}
\int dh p(h)h A(h)=-\Delta \int dh \frac{dp(h)}{dh}A(h)=\Delta \int dh p(h)\frac{\partial A(h)}{\partial h},
\end{align}
one finds
\begin{align}
\label{Eq_sh_conn}
\overline{<s_i>h_j}=\overline{\frac{\partial W[\mathbf{H},\mathbf{\hat{H}};\mathbf{h}]}{\partial \hat{H}_i} h_j}
=\Delta \; \overline{\frac{\partial W[\mathbf{H},\mathbf{\hat{H}}; \mathbf{h}]}{\partial \hat{H}_i\partial H_j}} = \Delta \; \hat{G}_{c,ij} \ ,
\end{align}
where
\begin{align}
\hat{G}_{c,ij}= \frac{\partial W_1[\mathbf{H},\mathbf{\hat{H}}]}{\partial \hat{H}_i \partial H_j}.
\end{align}
When all the sources are uniform, this yields
\begin{align}
\label{suscep}
\frac{\partial m(H,\hat{H})}{\partial H}=\hat{G}_{c}({\bf k=0})=\frac{1}{\Delta}\, \frac{1}{N}\sum_{i,j}\overline {<s_i >h_j}  \ .
\end{align}
where $\hat{G}_{c}({\bf k})$ is the Fourier transform of $\hat{G}_{c,ij}$. (In particular, this equation is valid in the limit $\hat{H} \rightarrow \pm \infty$, {\it i.e.} along the hysteresis loop: surprisingly, it  seems that this extension of the `susceptibily sum-rule' to the nonequilibrium magnetization curve has not been noticed before.)

To compute the average over disorder, the common procedure is to replicate the system $n$ times and take the limit $n \rightarrow 0$ at the end. (This is in contrast with the Martin-Siggia-Rose formalism in which the partition function is directly averaged over disorder, making the use of replicas unnecessary\cite{DS1996}; here indeed, the partition function in Eq.~(\ref{Eq4a}) is nontrivial so that one must average $\ln {\cal Z}$ and not simply $\cal Z$.) If one is interested in computing the cumulants of the random free-energy for generic arguments, one must introduce sources that act separately on each replica\cite{LW2003,TT2008,MT2010}.  After performing the average over the random-field distribution, we obtain a `replica partition function', 
\begin{equation}
\label{Eq7}
{\cal Z}_{rep}[\{\mathbf{H}^a,\mathbf{\hat{H}}^a\}]= \int \prod_{a=1}^{n} {\cal D}s^a{\cal D}\hat{s}^a\ e^{-S_{rep}[\{\mathbf{s}^a,\hat{\mathbf{s}}^a\}]+\sum_a[\mathbf{\hat{H}}^a.\mathbf{s}^a+\mathbf{H}^a.\hat{\mathbf{s}}^a]} 
\end{equation}
with the replicated action given by
\begin{align}
\label{Eq8}
S_{rep}[\{\mathbf{s}^a,\hat{\mathbf{s}}^a\}]= \sum_i \sum_{a}\hat{s}_{i}^{a}[V'(s_{i}^{a})-J\sum_{j/i}s_{j}^{a}]-\frac{\Delta}{2}\sum_i \sum_{a,b}\hat{s}_{i}^{a}\hat{s}_{i}^{b} \ .
\end{align}
The `thermodynamic potential' $W_{rep}[\{\mathbf{H}^a,\mathbf{\hat{H}}^a\}]=\ln {\cal Z}_{rep}[\{\mathbf{H}^a,\mathbf{\hat{H}}^a\}]$ can then be expanded in increasing number of free replica sums\cite{LW2003,TT2008,MT2010},
\begin{equation}
\label{Eq_free_replica_W}
W_{rep}[\{\mathbf{H}^a,\mathbf{\hat{H}}^a\}]=\sum_{p=1}^{\infty} \frac{1}{p!} \sum_{a_1,a_2,..,a_p}W_p[\mathbf{H}^{a_1}, \mathbf{\hat{H}}^{a_1};\mathbf{H}^{a_2}, \mathbf{\hat{H}}^{a_2};...;\mathbf{H}^{a_p}, \mathbf{\hat{H}}^{a_p}]
\end{equation}
where the $W_p$'s are continuous and symmetric functions of their arguments. This coincides with the cumulant expansion\cite{TT2008}. The thermodynamic potential $W_{rep}[\{\mathbf{H}^a,\mathbf{\hat{H}}^a\}]$ generates the `magnetizations' $m_{i}^{a}$ and $\hat{m}_{i}^{a}$ 
\begin{align}
\label{Eq9}
\frac{\partial W_{rep}}{\partial \hat{H}_{i}^{a}}&=<s_{i}^{a}>=m_{i}^{a}\nonumber\\
\frac{\partial W_{rep}}{\partial H_{i}^{a}}&=<\hat{s}_{i}^{a}>=\hat{m}_{i}^{a} \ ,
\end{align}
and the correlation (or Green's) functions, \textit{e.g.} at the pair level,
\begin{align}
\label{Eq10}
G_{ij}^{ab}&=\frac{\partial W_{rep}}{\partial \hat{H}_{i}^{a}\partial \hat{H}_{j}^{b}}=<s_{i}^{a}s_{j}^{b}>-<s_{i}^{a}><s_{j}^{b}>\nonumber\\
\hat{G}_{ij}^{ab}&=\frac{\partial W_{rep}}{\partial H_{i,a}\partial \hat{H}_{j,b}}=<s_{i}^{a}\hat{s}_{j}^{b}>-<s_{i}^{a}><\hat{s}_{j}^{b}>\nonumber\\
\hat{\hat{G}}_{ij}^{ab}&=\frac{\partial W_{rep}}{\partial H_{i}^{a}\partial H_{j}^{b}}=<\hat{\hat{s}}_{i}^{a}\hat{\hat{s}}_{j}^{b}>-<\hat{\hat{s}}_{i}^{a}><\hat{\hat{s}}_{j}^{b}> \ ,
\end{align}
where $<...>$ denotes an average over the replicated action. 
As $W_{rep}[\{\mathbf{H}^a,\mathbf{\hat{H}}^a\}]$ above, the magnetizations and the correlation functions can be expanded in increasing number of free replica sums:
\begin{equation}
\label{Eq_free_replica_m}
m_{i,a}[\{\mathbf{H}^e,\mathbf{\hat{H}}^e\}]=m_{i}^{[0]}[\mathbf{H}^a,\mathbf{\hat{H}}^a]+\sum_{e}m_{i}^{[1]}[\mathbf{H}^a, \mathbf{\hat{H}}^a \vert \mathbf{H}^{e}, \mathbf{\hat{H}}^{e}] + \frac{1}{2} \sum_{e,f} m_i^{[2]}[\mathbf{H}^a, \mathbf{\hat{H}}^a \vert \mathbf{H}^{e}, \mathbf{\hat{H}}^{e};\mathbf{H}^{f}, \mathbf{\hat{H}}^{f}] + ... ,
\end{equation}
and similarly for $\hat{m}_{i}^{a}$, whereas after  decomposing the two-point functions as  $G_{ij}^{ab}=G_{c,ij}^{a}\delta_{ab}+G_{d,ij}^{ab}$ (where $G_{d,ij}^{ab}$ does not contain any explicit Kronecker delta), one has \cite{TT2008, MT2010}
\begin{equation}
\label{Eq_free_replica_Gc}
G_{c,ij}^{a}[\{\mathbf{H}^e,\mathbf{\hat{H}}^e\}]=G_{c,ij}^{[0]}[\mathbf{H}^a, \mathbf{\hat{H}}^a]+\sum_{e}G_{c,ij}^{[1]}[\mathbf{H}^a, \mathbf{\hat{H}}^a \vert \mathbf{H}^{e}, \mathbf{\hat{H}}^{e}] + \frac{1}{2} \sum_{e,f} G_{c,ij}^{[2]}[\mathbf{H}^a, \mathbf{\hat{H}}^a \vert \mathbf{H}^{e}, \mathbf{\hat{H}}^{e};\mathbf{H}^{f}, \mathbf{\hat{H}}^{f}] + ...,
\end{equation}
\begin{align}
\label{Eq_free_replica_Gd}
G_{d,ij}^{ab}[\{\mathbf{H}^e,\mathbf{\hat{H}}^e\}]&=G_{d,ij}^{[0]}[\mathbf{H}^a, \mathbf{\hat{H}}^a;\mathbf{H}^b, \mathbf{\hat{H}}^b] +\sum_{e}G_{d,ij}^{[1]}[\mathbf{H}^a, \mathbf{\hat{H}}^a;\mathbf{H}^b, \mathbf{\hat{H}}^b \vert \mathbf{H}^{e}, \mathbf{\hat{H}}^{e}] \nonumber\\
&+ \frac{1}{2} \sum_{e,f} G_{d,ij}^{[2]}[\mathbf{H}^a, \mathbf{\hat{H}}^a;\mathbf{H}^b, \mathbf{\hat{H}}^b \vert \mathbf{H}^{e}, \mathbf{\hat{H}}^{e};\mathbf{H}^{f}, \mathbf{\hat{H}}^{f}] + ... ,
\end{align}
and similarly for $\hat{G}_{c,ij}^a,\hat{G}_{d,ij}^{ab}$ and $\hat{\hat{G}}_{c,ij}^a,\hat{\hat{G}}_{d,ij}^{ab}$.

In the present approach, however, the central object is not $W_{rep}$ but the  `effective action'  $\Gamma_{rep}$ which is the Legendre transform of $W_{rep}$ with respect to the two sets of sources $\{\mathbf{H}^a\}$ and $\{\mathbf{\hat{H}}^a\}$, 
\begin{align}
\label{Eq11}
\Gamma_{rep}[\{\mathbf{m}^a,\hat{\mathbf{m}}^a\}]=-W_{rep}[\{\mathbf{H}^a,\mathbf{\hat{H}}^a\}]+\sum_{a}\big (\mathbf{m}^a.\mathbf{\hat{H}}^a+\hat{\mathbf{m}}^a.\mathbf{H}^a \big )  \ ,
\end{align}
so that 
\begin{align}
\label{Eq12}
H_{i}^{a}&=\frac{\partial  \Gamma_{rep}}{\partial \hat{m}_{i}^{a}}\nonumber\\
\hat{H}_{i}^{a}&=\frac{\partial \Gamma_{rep}}{\partial m_{i}^{a}} \ .
\end{align}

$\Gamma_{rep}$ is the generating functional of the `direct' correlation functions [or one-particle irreducible (1PI) functions, or else  proper vertices, in field-theoretic language]. At the pair level,
\begin{align}
C_{ij}^{ab}&=\frac{\partial \Gamma_{rep}}{\partial m_{i}^{a}\partial m_{j}^{b}} \nonumber\\
\hat{C}_{ij}^{ab}&=\frac{\partial \Gamma_{rep}}{\partial m_{i}^{a}\partial \hat{m}_{j}^{b}}\nonumber\\
\hat{\hat{C}}_{ij}^{ab}&=\frac{\partial \Gamma_{rep}}{\partial \hat{m}_{i}^{a}\partial \hat{m}_{j}^{b}}\ .
\end{align}
As above, $\Gamma_{rep}$, $H_{i}^{a},\hat{H}_{i}^{a}$, $C_{ij}^{ab}$, etc... can be expanded in increasing number of free replica sums.

In the following, all the two-point functions will be put together as components of a matrix with both replica indices and spatial coordinates
\begin{align}
\label{Eq13a}
\underline{\underline{\mathbf{G}}}=\left(\begin{array}{c}\mathbf G \ \bf{\hat{G}}\\ 
\bf \hat{G} \ \bf{\hat{\hat{G}}}\end{array} \right) \ ,
 \end{align}
where $\bf G$, $\bf{\hat G}$, $\bf \hat{\hat G}$ have for elements $G^{ab}_{ij}$,  $\hat{G}^{ab}_{ij}$,  $\hat{\hat G}^{ab}_{ij}$; a similar notation is used for the direct correlation functions, all collected in $\underline{\underline{\bf C}}$. The matrix  $\underline{\underline{\bf{C}}}$ is then just the inverse of  $\underline{\underline{\bf{G}}}$, \textit{i.e.}
\begin{align}
\label{Eq13}
 \bf{\underline{\underline{C}}}= {\bf \underline{\underline{G}}}^{-1} \ .
\end{align}

The matrix inversion with respect to spatial coordinates is easily realized by Fourier transformation when all the sources are taken as uniform. The inversion with respect to the replica indices can be performed by using the expansion in free replica sums for ${\bf \underline{\underline{G}}}$ and ${\bf \underline{\underline{C}}}$ (see Eqs.~(\ref{Eq_free_replica_Gc},\ref{Eq_free_replica_Gd})) and proceeding to a term-by-term identification. The zeroth-order terms are then given by
\begin{align}
\label{Eq_zeroth_orderOZ}
&\mathbf{\underline{\underline{G}}}^{[0]}_c(\mathbf{k};  m^a,\hat{m}^a) =\mathbf{\underline{\underline{C}}}^{[0]}_c(\mathbf{k}; m^a,\hat{m}^a)^{-1}\\ &
\mathbf{\underline{\underline{G}}}^{[0]}_d(\mathbf{k}; m^a,\hat{m}^a; m^b,\hat{m}^b) =- \mathbf{\underline{\underline{C}}}^{[0]}_c(\mathbf{k}; m^a,\hat{m}^a)^{-1}\mathbf{\underline{\underline{C}}}^{[0]}_d(\mathbf{k}; m^a,\hat{m}^a; m^b,\hat{m}^b) \mathbf{\underline{\underline{C}}}^{[0]}_c(\mathbf{k}; m^b,\hat{m}^b)^{-1} \ ,
 \end{align}
where $\mathbf{\underline{\underline{G}}}^{[0]}_{c,d}$ are $2 \times 2$ matrices containing the components $\mathbf G_{c,d}$, $ \hat{{\bf G}}_{c,d}$, $\hat{\hat{{\bf G}}}_{c,d}$ as in Eq.~(\ref{Eq13a}), and similarly for $\mathbf{\underline{\underline{C}}}^{[0]}_{c,d}$. Note that the zeroth-order functions obtained when fixing the replica sources $\{H^a, \hat{H}^a\}$ and fixing the replica `magnetizations' $\{m^a, \hat{m}^a\}$ are generically related through $G(H^a, \hat{H}^a)\equiv G(m^{[0]}(H^a, \hat{H}^a), \hat{m}^{[0]}(H^a, \hat{H}^a))$ with $m^{[0]}$ and $\hat{m}^{[0]}$ being the zeroth-order expressions  as in Eq.~(28). Since the zeroth-order contributions already contain the physics of the problem, we will not consider higher-order terms and will drop the superscript $[0]$ in the following. 

When all replica sources or magnetizations are taken as equal, and provided that this limiting process is regular enough (this will be discussed in more detail below), replica symmetry is recovered and one obtains a set of `Ornstein-Zernike'  (OZ) equations (in the language of liquid-state theory\cite{HM1986}) which takes the following explicit form:
\begin{align}
\label{EqOZc}
G_c(\mathbf{k})&=\frac{\hat{\hat{C}}_c(\mathbf{k})}{C_c(\mathbf{k})\hat{\hat{C}}_c(\mathbf{k})-\hat{C}_c(\mathbf{k})^2}\nonumber\\
\hat{G}_c(\mathbf{k})&=-\frac{\hat{C}_c(\mathbf{k})}{C_c(\mathbf{k})\hat{\hat{C}}_c(\mathbf{k})-\hat{C}_c(\mathbf{k})^2}\nonumber\\
\hat{\hat{G}}_c(\mathbf{k})&=\frac{C_c(\mathbf{k})}{C_c(\mathbf{k})\hat{\hat{C}}_c(\mathbf{k})-\hat{C}_c(\mathbf{k})^2}
\end{align}
and
\begin{align}
\label{EqOZd}
G_d(\mathbf{k})&=-[C_d(\mathbf{k})G_c(\mathbf{k})^2+2\hat{C}_d(\mathbf{k})G_c(\mathbf{k})\hat{G}_c(\mathbf{k})+\hat{\hat{C}}_d(\mathbf{k})\hat{G}_c(\mathbf{k})^2]\nonumber\\
\hat{G}_d(\mathbf{k})&=-[C_d(\mathbf{k})G_c(\mathbf{k})\hat{G}_c(\mathbf{k})+\hat{C}_d(\mathbf{k})(G_c(\mathbf{k})\hat{\hat{G}}_c(\mathbf{k})+\hat{G}_c(\mathbf{k})^2) +\hat{\hat{C}}_d(\mathbf{k})\hat{\hat{G}}_c(\mathbf{k})\hat{C}_c(\mathbf{k})]\nonumber\\
\hat{\hat{G}}_d(\mathbf{k})&=-[\hat{\hat{C}}_d(\mathbf{k})\hat{\hat{G}}_c(\mathbf{k})^2+2\hat{C}_d(\mathbf{k})\hat{\hat{G}}_c(\mathbf{k})\hat{G}_c(\mathbf{k})+C_d(\mathbf{k})\hat{G}_c(\mathbf{k})^2] \ ,
\end{align}
where all functions depend on $m$ and $\hat m$.

When the replica sources are not equal, the zeroth-order terms have the capability to capture a nonanalytic behavior in the dependence on the replica sources or replica magnetizations. This important feature is already displayed by the noninteracting system corresponding to $J=0$ (hereafter called the `reference' system) whose properties are listed in the appendix (Eq.~ (\ref{EqZrep0}) and below with $\hat{K}_c=0$ and $\hat{\hat{K}}_d=\Delta$). For instance, the 2-replica function $G_{d,ij}(H^a,\hat{H};H^b,\hat{H})$ behaves like
 \begin{align}
\label{Gcusp}
G_{d,ij}(H^a,\hat{H}; H^b,\hat{H})=G_{d,ij}(H,\hat{H};H,\hat{H})+G_{d,ij}^{cusp}(H,\hat{H})\vert H^a-H^b\vert+ O((H^a-H^b)^2)
 \end{align}
when $H^a,H^b \rightarrow H$. We stress that this is not a spurious behavior due to the presence of a cusp in $V(s)$. Indeed, even if $V(s)$ were a smooth double-well potential, it would  be necessary to distinguish between the two minima in order to count the metastable states and, whatever the method, this would introduce a nonanalyticity in $W[\mathbf{H},\mathbf{\hat{H}};\mathbf{h}]$ and lead to a cusp in the 2-replica function $G_d$. This behavior is thus an intrinsic feature of the problem under consideration and is also intimately connected to the magnetization discontinuities along the hysteresis loop, \textit{i.e.} to the avalanches. A similar connection is  discussed in Ref.~\cite{LW2009} in the context of random elastic systems where the  statistics of static avalanches (or `shocks')  is studied via the functional renormalization group. (In this case, however, the cusp only shows up in the course of the renormalization flow whereas it is always present here, even in the large disorder regime.)

To show that the cusp is related to the (unnormalized) second moment of the avalanche size distribution, we consider the quantity
 \begin{align}
\label{EqDelta}
G_d(\mathbf{k=0};H^a;H^b)= N\left[ \overline {m(H^a;{\bf h})m(H^b;{\bf h})}- m(H^a)m(H^b)\right],
 \end{align}
where it is implicit that the limit $\hat{H}\rightarrow \pm \infty$ corresponding to the hysteresis loop (in the large-disorder regime) has been taken. Then, using Eqs. (\ref{magcusp}) and (\ref{Gcusp}), taking the second derivative with respect to $H^a$ and $H^b$ in the limit $H^a,H^b\rightarrow H$, and identifying the singular contributions proportional to $\delta(H^a-H^b)$ on both sides of the equation lead to
 \begin{align}
\label{S2moment}
N\, \overline{\sum_{\alpha}S^2_{\alpha}\delta(H-H_{\alpha})}=N \int dS\, S^2 \rho(S,H)= -2 \,G_d^{cusp}(\mathbf{k=0};H).
 \end{align}

The nonanalyticity in $W[\mathbf{H},\mathbf{\hat{H}};\mathbf{h}]$ also implies that the 2-replica correlation function $\hat{\hat{G}}_{d,ij}(H^a,\hat{H}^a; H^b,\hat{H}^b)$ contains a singular contribution proportional to $\delta(H^a-H^b)$. This induces singular contributions in the three disconnected direct correlation functions via the OZ equations and leads to formally diverging terms proportional to `$\delta(0)$'  when the replica fields are equal. The function $\hat{\hat{G}}_d$, however, has no obvious physical meaning and this problem is in principle harmless, although it may be a source of difficulties in an approximate treatment, as will be discussed below in section V.

Finally, it is instructive to examine the behavior of the correlation functions in the limit $\hat{H}\rightarrow \pm \infty$, \textit{i.e.} along the two branches of the hysteresis loop in the large-disorder regime. Assuming that the general behavior is the same as in the reference system (which seems quite reasonable, at least in the regime under consideration), we find
\begin{align}
G_{c} & =O(e^{-\vert\hat{H}\vert}),\ \ \ C_{c}=O(\hat{H})\nonumber\\
\hat{G}_{c} & =O(1), \ \ \ \ \ \ \hat{C}_{c}=O(1)\nonumber\\
\hat{\hat{G}}_{c} &=O(\hat{H}), \ \ \ \hat{\hat{C}}_{c}=O(e^{-\vert\hat{H}\vert})
\end{align}
and
\begin{align}
G_{d} & =O(1),\ \ \ \ \ C_{d}=\delta(0) O(\hat{H}^2)+O(\hat{H}^2)\nonumber\\
\hat{G}_{d} & =O(\hat{H}), \ \ \ \ \ \ \ \hat{C}_{d}=O(\hat{H})\nonumber\\
\hat{\hat{G}}_{d} &=\delta(0) O(\hat{H}^2)+O(\hat{H}^2), \ \ \ \hat{\hat{C}}_{d}=O(1) \ ,
\end{align}
where the notation indicates that both the regular and the singular contributions of $\hat{\hat{G}}_{d}$ and $C_{d}$ are of order $\hat{H}^2$. Note that we have considered the dependence on the source $\hat{H}$. A similar dependence is found on $\hat m$ when the latter goes to plus or minus infinity (recall that we work at the zeroth-order level of the expansion in free replica sums).

It is not surprising that $G_c$  vanishes as $\hat{H}\rightarrow \pm \infty$.  This is due to the already mentioned fact that there is only one metastable state along the loop at a given field so that the statistical fluctuations only come from the quenched disorder and are thus contained in the disconnected functions. (As a consequence, $\hat{\hat{C}}_c$ also vanishes.) The important feature is that  $G_c$ and $\hat{\hat{C}}_c$ vanish exponentially fast with $\hat{H}$. Indeed, this implies that  the OZ equations for the physically relevant functions $\hat{G}_c$ and $G_d$ take in this limit the much simpler form:
\begin{align}
\label{RPA1}
\hat{G}_c(\mathbf{k})&=\frac{1}{\hat{C}_c(\mathbf{k})}\nonumber\\
G_d(\mathbf{k})&=-\frac{\hat{\hat{C}}_d(\mathbf{k})}{\hat{C}_c(\mathbf{k})^2}  \ .
\end{align}
Note that $\hat{\hat{C}}_d(\mathbf{k})$ does not contain any diverging part when $\hat{H}\rightarrow \pm \infty$ (or $\hat{m}\rightarrow \pm \infty$), as it must be.

\section{Random phase approximation (RPA)}

Equipped with the formalism of the previous section, we wish to introduce approximations based on assumptions on the  structure of the direct correlation functions (proper vertices).  Formally, these functions may be written as
\begin{align}
{\bf \underline{\underline{C}}}={\bf \underline{\underline{C}}}^{(0)}-\bf {\underline{\underline{\Sigma}}}
\end{align}
where ${\bf \underline{\underline{C}}}^{(0)}$ is the `bare' inverse propagator matrix (not taking into account the local potential $V$), whose components satisfy
\begin{align}
\label{EqSD}
C^{(0)ab}_{ij}&=0\nonumber\\
\hat{C}^{(0)ab}_{ij}&=-J \lambda_{ij} \, \delta_{ab}\nonumber\\
\hat{\hat{C}}^{(0)ab}_{ij}&=-\Delta \ ,
\end{align}
and $\bf {\underline{\underline{\Sigma}}}$ is a `self-energy'  matrix (the minus sign is chosen so to match the usual definition of this quantity  in field theory and many-body physics). In the above equations, $\lambda_{ij}$ is $1$ if $i$ and $j$ are nearest neighbors on the lattice, zero otherwise.

From now on in this section, except if explicitly stated, we only consider uniform sources that are equal for all replicas. Our first approximation, akin to the usual random phase approximation (RPA), consists in assuming that the self-energies are identical to those of the noninteracting ($J=0$) reference system.  In Fourier space, this then leads to
\begin{align}
\label{EqCRPA}
C_c^{RPA}(\mathbf{k})&=C_c^{ref}, \ \ \ \ \ \ \ \ \ \ \ \ \ \ \ \ \ C_d(\mathbf{k})=C_d^{ref}\nonumber\\
\hat{C}_c^{RPA}(\mathbf{k})&=\hat{C}_c^{ref}-qJ\lambda({\bf k}),\ \ \ \ \hat{C}_d(\mathbf{k})=\hat{C}_d^{ref}\nonumber\\
\hat{\hat{C}}_c^{RPA}(\mathbf{k})&=\hat{\hat{C}}_c^{ref}, \ \ \ \ \ \ \ \ \ \ \ \ \ \ \ \ \ \hat{\hat{C}}_d(\mathbf{k})=\hat{\hat{C}}_d^{ref}
\end{align}
where $q=2d$ is the connectivity of the hypercubic lattice and $\lambda({\bf k})=(1/q) \sum_{{\bf e}}\exp(i {\bf k}.{\bf e})$ its characteristic function. The direct correlation functions of the reference system are obtained from the Green's functions computed in the appendix  [Eqs. (\ref{EqGc})-(\ref{EqhhGc}) and (\ref{EqDeltacusp})-(\ref{Gdasymp}) with $\hat{K}_c=0$ and $\hat{\hat{K}}_d=\Delta$]. The RPA effective action is immediately obtained by integrating the `susceptibility sum-rule'  $\partial^2 (\Gamma/N)/\partial m\partial \hat{m}=\hat{G}_c^{-1}(\mathbf{k=0})=\hat{C}_c(\mathbf{k=0})$, which yields
\begin{align}
\frac{1}{N}\Gamma^{RPA}(m,\hat{m})=\frac{1}{N}\Gamma^{ref}(m,\hat{m})-qJm\hat{m}.
\end{align}
From this expression, one obtains
\begin{align}
\label{Eqfield}
H(m,\hat{m})&=H^{ref}(m,\hat{m})-qJm \nonumber\\
\hat{H}(m,\hat{m})&=\hat{H}^{ref}(m,\hat{m})-qJ\hat{m} \ .
\end{align}
It can be easily checked that  $\Sigma_Q^{RPA}(m,H)/N=-\Gamma^{RPA}(m,\hat{m})/N+\hat{m}H(m,\hat{m})$ is identical to the quenched complexity $\Sigma_Q$ (not to be confused with a self-energy) obtained in the mean-field model of Ref.~\cite{RM2009} (with $qJ$ replaced by $J$). (On the other hand, the auxiliary field $\hat{H}$ does not coincide with the parameter $g$ introduced in this reference. In particular, $\hat{H}$ satisfies the Legendre relation $\partial (\Sigma_Q(m,H)/N)/\partial m=-\hat{H}$, in contrast with $g$. In this respect, the RPA is a nicer way of obtaining the mean-field limit.)
\begin{figure}[hbt]
\begin{center}
\includegraphics[width=9cm]{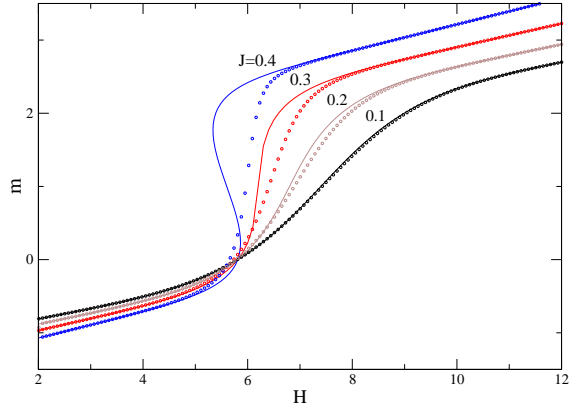}
\caption{\label{Fig2} Magnetization along the ascending branch of the hysteresis loop: comparison of the predictions of the random-phase approximation (solid lines)  to simulation data (circles) for various values of the coupling $J$ (with $k=8$ and $\Delta=4$). The simulation data are averaged over $1000$ disorder realizations of linear size $L=30$.}
\end{center}
\end{figure}

One expects the RPA to be valid when the coupling $J$ is sufficiently weak, or, equivalently, when $\Delta$ and $k$ are sufficiently large (all `thermodynamic' quantities can be expressed in terms of the reduced variables $\sqrt{\Delta}/J$, $k/J$, and $H/J$). This is indeed what is observed in Fig. \ref{Fig2} where  the predictions of the RPA for the magnetization along the ascending branch of the hysteresis loop are compared to numerical simulations (the descending branch is obtained using the symmetry $H\rightarrow -H$, $m\rightarrow -m$). The theoretical value  is given by  $m^{RPA}(H)=m^{ref}(H^{ref})$ where $m^{ref}$ is given by Eq.~(\ref{Eqmag}) with $\hat{K}_c=0$ and $\hat{\hat{K}}_d=\Delta$, and from the above Eq. (\ref{Eqfield}) the ``displaced'' field $H^{ref}$ is solution of the implicit equation $H=H^{ref}-qJm^{ref}(H^{ref})$. One can see that the agreement is very good for $J=0.1$ and deteriorates as $J$ increases. In particular, the RPA overestimates the slope $\partial m/\partial H$ and the magnetization curve already exhibits a reentrant behavior for $J=0.4$  (typical of a mean-field theory below the critical point) whereas the actual system is still in the large-disorder regime.

\begin{figure}[hbt]
\begin{center}
\includegraphics[width=9cm]{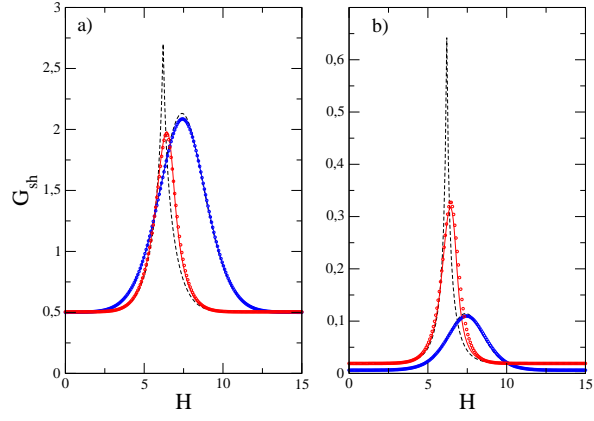}
\caption{\label{Fig3} Spin-random field correlation function $G_{sh}({\bf r})$ at ${\bf r=0}$ (a) and ${\bf r=e}$ (b) along the ascending branch of the hysteresis loop for $J=0.1$ (blue) and $J=0.3$ (red). The predictions of the RPA (dashed lines) and the LSEA (solid lines) are compared to the simulation data (circles).}
\end{center}
\end{figure}

\begin{figure}[hbt]
\begin{center}
\includegraphics[width=9cm]{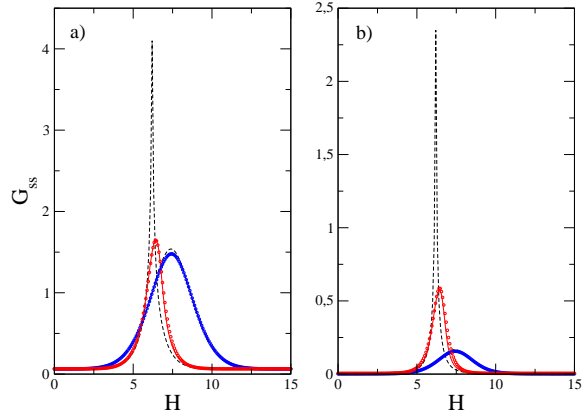}
\caption{\label{Fig4} Same as Fig. 3 for the spin-spin correlation function $G_{ss}({\bf r})$.}
\end{center}
\end{figure}

The spin-spin and spin-random-field correlation functions along the loop for $\mathbf{r=0}$ and $\mathbf{r=e}$ are shown in Figs. \ref{Fig3} and \ref{Fig4}.  They are  defined by (see section III)
\begin{align}
\label{EqGss}
G_{ss}(\mathbf{r})\equiv \overline{s_0 s_{\mathbf{r}}}-m^2=\lim_{\hat{H}\rightarrow \pm \infty}G_d(\mathbf{r})
\end{align}
and
\begin{align}
\label{EqGsh}
G_{sh}(\mathbf{r})\equiv \overline{s_0 h_{\mathbf{r}}}=\Delta\lim_{\hat{H}\rightarrow \pm \infty}\hat{G}_c(\mathbf{r})\  .
\end{align}
Eqs.~(\ref{RPA1}) and (\ref{EqCRPA}) then yield
\begin{align}
G_{sh}^{RPA}(\mathbf{r})=G_{sh}^{ref}P(\mathbf{r};z)
\end{align}
\begin{align}
G_{ss}^{RPA}(\mathbf{r})=G_{ss}^{ref}[P(\mathbf{r};z)+z P'(\mathbf{r};z)]
\end{align}
where 
\begin{align}
z=z^{RPA}=\frac{qJ}{\hat{C}_c^{ref}} 
\end{align}
and
\begin{align}
P({\bf r};z)=\frac{1}{(2\pi)^3}\int_{-\pi}^{\pi}d^3{\bf k}\frac{e^{i{\bf k}.{\bf r}}}{1-z\lambda({\bf k})} 
\end{align}
is the lattice Green function; in addition, $P'(\mathbf{r};z)=dP(\mathbf{r};z)/dz$ ($z$ and the correlation functions of the reference system are functions of $m$ or $H$).  One can see that the RPA correctly predicts the values of the correlation functions along the hysteresis loop when $J$ is small (we have checked that the agreement with the simulations remains good at the second and third nearest-neighbor distances) but considerably overestimates these values when $J$ increases and the slope $\partial m/\partial H$ becomes large (for instance around $H\approx 6.45$ for $J=0.3$). This can be traced back to an overestimation of the value of $z$ (note that the RPA susceptibility diverges for $z=1$).


Finally, it is also interesting to examine the RPA correlation functions for $\hat {H}$ finite, which corresponds to metastable states {\it inside} the hysteresis loop in the $H-m$ plane. For conciseness, we only consider the connected functions, as given by Eqs. \ref{EqOZc}. We then write
 \begin{align}
\label{RPA3}
C_c^{RPA}(\mathbf{k})\hat{\hat{C}}_c^{RPA}(\mathbf{k})-\hat{C}_c^{RPA}(\mathbf{k})^{2}&=C_c^{ref}\hat{\hat{C}}_c^{ref}-[\hat{C}_c^{ref}-qJ\lambda(\mathbf{k}))]^2\nonumber\\
&=[C_c^{ref}\hat{\hat{C}}_c^{ref}-(\hat{C}_c^{ref})^2][1-z_1\lambda(\mathbf{k})][1-z_2\lambda(\mathbf{k})]
\end{align}
 where
 \begin{align}
 z_{1,2}=z_{1,2}^{RPA}=\frac{qJ}{\hat{C}_c^{ref}\pm \sqrt{C_c^{ref}\hat{\hat{C}}_c^{ref}}} \ ,
\end{align}
which yields
\begin{align}
G_c^{RPA}(\mathbf{r})&=G_c^{ref}\big[\frac{z_1}{z_1-z_2}P(\mathbf{r};z_1)-\frac{z_2}{z_1-z_2}P(\mathbf{r};z_2)\big]\nonumber\\
\hat{G}_c^{RPA}(\mathbf{r})&=\hat{G}_c^{ref}\big[\frac{z_1-z}{z_1-z_2}P(\mathbf{r};z_1)-\frac{z_2-z}{z_1-z_2}P(\mathbf{r};z_2)\big]\nonumber\\
\hat{\hat{G}}_c^{RPA}(\mathbf{r})&=\hat{\hat{G}}_c^{ref}\big[\frac{z_1}{z_1-z_2}P(\mathbf{r};z_1)-\frac{z_2}{z_1-z_2}P(\mathbf{r};z_2)\big]
\end{align}
where $z=z^{RPA}$ as defined above.

\begin{figure}[hbt]
\begin{center}
\includegraphics[width=7.5cm]{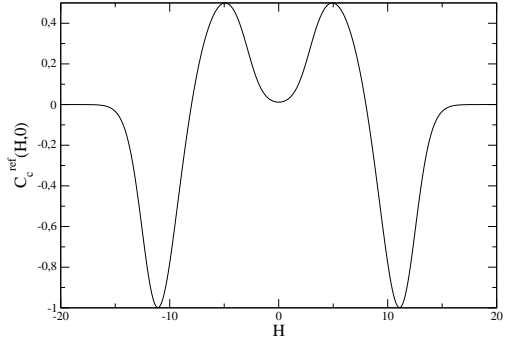}
\caption{\label{Fig5} Direct correlation function $C_c^{ref}(H,\hat{H})$ in the reference system ($J=0$) as a function of $H$ for $\hat{H}=0$.}
\end{center}
\end{figure}

\begin{figure}[hbt]
\begin{center}
\includegraphics[width=7.5cm]{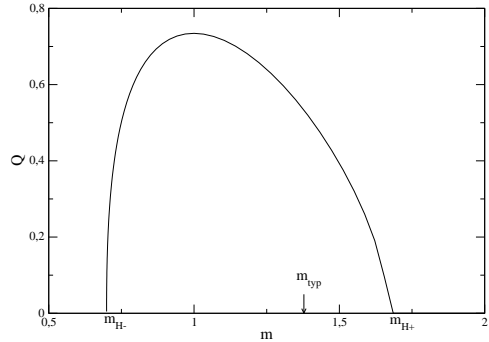}
\caption{\label{Fig6} Wave-vector $Q$ as a function of the magnetization $m(\hat{H})$ of the metastable states for $J=0.1$ and $H=7$. $Q$ is non-zero for $m_{typ}\equiv m(\hat{H}=0)$, the magnetization of the typical states, but zero on the hysteresis loop (as $\hat{H} \rightarrow \pm \infty$).}
\end{center}
\end{figure}

It turns out that $\hat{C}^{ref}_c$ is a positive function of $H$ and $\hat{H}$ and $\hat{\hat{C}}^{ref}_c$ is negative, whereas the sign of $C^{ref}_c$ may change, as shown in Fig.~\ref{Fig5} for $\hat{H}=0$. Therefore, depending on whether $C^{ref}_c$ is negative or positive, $z_1$ and $z_2$ are real or complex conjugates, respectively.  However, one can check that the correlation functions are always real, as it must be. On the other hand their behavior changes with the sign of $C^{ref}_c$, as illustrated by the leading asymptotic behavior as $r \rightarrow \infty$. For instance, using the asymptotic expansion of $P(n,n,n;z)$ \cite{J2004}, one finds that
\begin{align}
\hat{G}_c^{RPA}(n,n,n)\sim\hat{G}_c^{ref}\frac{\sqrt{3}}{2\pi n}\big[\frac{z_1-z}{z_1-z_2}e^{-\lambda_1 n}-\frac{z_2-z}{z_1-z_2}e^{-\lambda_2 n}\big]\nonumber\\
\end{align}
where $\lambda_{1,2}=3\ln [z_{1,2}/[1-\sqrt{1-z_{1,2}^2}]$. Therefore, when $C^{ref}_c>0$ and $z_{1,2}$ are complex conjugates, the typical Ornstein-Zernike fall-off $e^{-\xi n}/n$  is modulated by oscillations at a wavevector $Q=n\Im(\lambda_{1,2})$ where $\Im$ denotes the imaginary part. For a given value of the magnetic field $H$, both the correlation length and the wavevector $Q$ depend continuously on $\hat{H}$ (or, alternatively, on  the magnetization $m(\hat{H})$ of the metastable states), as shown in Fig.~\ref{Fig6}. This defines a `disorder line' in the $H-m$ plane where the qualitative behavior of the correlation function changes. In the case presented in Fig.~\ref{Fig6} (for a small value of $J$ for which the RPA is expected to be valid), one finds that $Q$ is non-zero for $\hat{H}=0$, that is for a typical metastable state at the field $H$. On the other hand, $Q$ is always zero for $\hat{H} \rightarrow \pm \infty$, that is along the hysteresis loop.

\section{Beyond the RPA }

\subsection{Local self-energy approximation (LSEA)}

The RPA is exact in the limit of infinite dimension or infinite lattice coordination but, as we have just seen, is not a good approximation for $d=3$ outside the  weak-coupling regime. To go beyond this regime, we shall assume that the fluctuations renormalize the Green's functions {\it without} changing the functional form of their spatial dependence. This amounts to assuming that the corresponding self-energies are purely local. This assumption is the starting point of an approximate theory similar to the so-called Optimized Random Phase Approximation (ORPA) in liquid-state theory\cite{HM1986} (see also\cite{KRT1999}), the `locator' approximation used in spin and Coulomb glasses\cite{FT1979,MI2004,MP2007}, and to the dynamical mean-field theory (DMFT) for quantum problems\cite{GKKR1996}. This type of approximation is known to be efficient when the physics is dominated by strong short-wavelength fluctuations rather than by long-wavelength fluctuations\cite{GKKR1996}. Accordingly, it cannot properly describe the vicinity of critical points but may capture for instance hysteresis and metastable effects that are already generated by the local potential in Eq.~(\ref{Eq2}). Note that the local self-energy approximation reduces to the RPA when $d\rightarrow \infty$, but may be expected to provide reasonable results even in $d=3$.

From the point of view of thermodynamic consistency, it is useful to formulate the theory at the level of the effective action within the two-particle irreducible (2PI) formalism where $\Gamma_{rep}$ is considered as a functional of both the magnetizations $\{m_{i}^{a},\hat{m}_{i}^{a}\} $ and the Green's functions\cite{CJT1974} (this is known as the `entropy functional' in classical systems\cite{P1990}).  Up to an additive constant, $\Gamma_{rep}$ can be written as
\begin{align}
\label{Eqfunc}
\Gamma_{rep}[\{\mathbf{m}^{a},\mathbf{\hat{m}}^{a}\},\mathbf{\underline{\underline{G}}}]
=S_{rep}[\{\mathbf{m}^{a},\mathbf{\hat{m}}^{a}\}]+\frac{1}{2} \mbox{Tr}\ln {\bf \underline{\underline{G}}}^{-1}+\frac{1}{2}\mbox{Tr}\,{\bf \underline{\underline{C}}}^{(0)}{\bf \underline{\underline{G}}}+\Phi_{rep}[\{\mathbf{m}^{a},\mathbf{\hat{m}}^{a}\},\mathbf{\underline{\underline{G}}}],
\end{align}
where $\mathbf{\underline{\underline{G}}}$ is defined in Eq.~(\ref{Eq13a}) and $\Phi_{rep}$ is the so-called Luttinger-Ward functional\cite{LW1960,BK1961}. ($\Phi_{rep}$ is also called the 2PI functional as it is in general  the sum of all two-particle irreducible diagrams built with the fully dressed Green's functions. In the present case, the singular nature of the local potential makes the definition of the vertices appearing in the diagrams tricky and Eq.~(\ref{Eqfunc}) must thus be considered as a nonperturbative definition of $\Phi_{rep}$.) In the above expression, the trace involves a sum over both replica and spatial indices. The crucial point is that the explicit dependence of $\Gamma_{rep}$ on the pair interactions is contained in the classical action $S_{rep}[\{\mathbf{m}^{a},\mathbf{\hat{m}}^{a}\}]$ and in the bare inverse propagator ${\bf \underline{\underline{C}}}^{(0)}$. On the other hand, for a given on-site potential, the functional $\Phi_{rep}$ is universal. By construction, the self-energies are obtained as
\begin{align}
\label{Eq_selfenergies}
\Sigma_{ij}^{ab}=-2\ \frac{\partial \Phi_{rep}}{\partial G_{ij}^{ab}},
\end{align}
etc...,  and the stationnarity of the functional $\Gamma_{rep}$ against the variations of the Green's functions provides the Schwinger-Dyson equations,
\begin{align}
\label{Eq_schwinger-dyson}
\mathbf{\underline{\underline{G}}}^{-1}= \mathbf{\underline{\underline{C}}}^{(0)} - \underline{\underline{\mathbf{\Sigma}}}
\end{align}
where one also has from Eq.~(\ref{Eq13}) that $\mathbf{\underline{\underline{G}}}^{-1}= \mathbf{\underline{\underline{C}}}$. At the extremum, $\Gamma_{rep}$ then identifies with the physical (1PI) replicated effective action defined in Eq.~(\ref{Eq11}).

At the level of $\Gamma_{rep}$, the  local  approximation consists in replacing $\Phi_{rep}$ by a sum of purely local contributions, $\sum_i \phi_{rep} (\{m_{i}^{a},\hat{m}_{i}^{a}\},\mathbf{\underline{\underline{G}}}_{ii})$. This readily implies that the self-energies are purely local. More specifically, restricting ourselves to the translationally invariant situation (uniform configurations), we find in Fourier space
\begin{align}
\label{Eq_SD_fourier}
\mathbf{\underline{\underline{C}}}^{ab}(\mathbf{k})= \mathbf{\underline{\underline{C}}}^{(0)ab}(\mathbf{k})-\mathbf{\underline{\underline{\Sigma}}}^{ab} \ ,
\end{align}
which expresses in a matrix form the equations for the components $C^{ab}, \hat C^{ab}, \hat{\hat C}^{ab}$ in terms of their counterparts in $\mathbf{\underline{\underline{C}}}^{(0)ab}$ and in the self-energy. According to Eq.~(\ref{EqSD}), the only nonzero components of $\mathbf{\underline{\underline{C}}}^{(0)ab}$ are $\hat{C}^{(0)ab}({\bf k})=  -qJ\lambda({\bf k})\delta_{ab}$ and $\hat{\hat C}^{(0)ab}({\bf k})=- \Delta$. The self-energies are then functions of $\{m^a,\hat{m}^a\}, J,k$ and $\Delta$ to be determined. 

As discussed for instance in Ref.~\cite{GKKR1996}, the simplest strategy for computing the above functions is to define a single-site effective action (the so-called  `impurity' model in strongly correlated Fermi systems) that involves the original on-site interaction and arbitrary quadratic terms. This model is in general exactly solvable and self-consistency equations are then obtained by imposing that the Green's functions of the single-site action coincide with the site-diagonal Green's functions of the original lattice model.

We thus introduce the single-site action
\begin{align}
\label{Sloc}
S_{0,rep}[\{s^a,\hat{s}^a\}]= -\frac{1}{2}\sum_{a,b}\left[K^{ab}s^a s^b+\hat{K}^{ab}(s^a \hat{s}^b+\hat{s}^a s^b)+\hat{\hat{K}}^{ab} \hat{s}^a\hat{s}^b\right]+\sum_{a} \hat{s}^{a}V'(s^{a})
\end{align}
and add external sources $H_0^{a},\hat{H}_0^{a}$ that fix  the same `magnetizations' $\hat{m}^a$ and $m^a$, respectively, as in the fully interacting system. The so-called `Weiss fields' $K^{ab},\hat{K}^{ab},\hat{\hat{K}}^{ab}$  must be chosen so that the Green's functions $G_0^{ab}, \hat{G}_0^{ab}, \hat{\hat{G}}_0^{ab}$ of the effective single-site model coincide with the on-site Green's functions $G_{ii}^{ab}, \hat{G}_{ii}^{ab}, \hat{\hat{G}}_{ii}^{ab}$ of the original model, \textit{i.e.}
\begin{align}
\label{Eq_green_local}
\mathbf{\underline{\underline{G}}}^{ab}(\mathbf r=\mathbf 0)=\mathbf{\underline{\underline{G}}}_{\:0}^{ab}, 
\end{align}
with identical self-energies $\Sigma^{ab}_{ij}=\Sigma^{ab}\delta_{ij}$, $\hat{\Sigma}^{ab}_{ij}=\hat{\Sigma}^{ab}\delta_{ij}$, $\hat{\hat{\Sigma}}^{ab}_{ij}=\hat{\hat{\Sigma}}^{ab}\delta_{ij}$\cite{note3}.  By definition of the self-energies, the direct correlation functions of the effective model are given by
\begin{align}
\label{Eq_schwinger-dyson_0}
\mathbf{\underline{\underline{C}}}_{\:0}^{ab}= - \mathbf{\underline{\underline{K}}}^{ab} - \mathbf{\underline{\underline{\Sigma}}}^{ab}
\end{align}
as $\mathbf{\underline{\underline{C}}}^{(0)ab}_{\:0}= - \mathbf{\underline{\underline{K}}}^{ab}$ in the single-site model, with the matrix $\mathbf{\underline{\underline{K}}}^{ab}$ collecting the Weiss fields $K^{ab},\hat{K}^{ab},\hat{\hat{K}}^{ab}$. Subtracting from Eq. (\ref{Eq_SD_fourier}) then yields
\begin{align}
\label{EqLSEA}
\mathbf{\underline{\underline{C}}}^{ab}(\mathbf{k})=\mathbf{\underline{\underline{C}}}^{(0)ab}(\mathbf{k})+ \mathbf{\underline{\underline{K}}}^{ab}+\mathbf{\underline{\underline{C}}}_{\:0}^{ab}  
\end{align}
and the consistency requirement between $\mathbf{\underline{\underline{G}}}_{\:0}^{ab}$ and $\mathbf{\underline{\underline{G}}}_{\:ii}^{ab}$ provides the following equation,
\begin{align}
\label{Eq_self-consist_DMFT}
\mathbf{\underline{\underline{G}}}_{\:0}=\mathbf{\underline{\underline{C}}}_{\:0}^{-1}= \int d{\bf k} \left[\mathbf{\underline{\underline{C}}}^{(0)}(\mathbf k)+ \mathbf{\underline{\underline K}} +\mathbf{\underline{\underline{C}}}_{\:0}\right] ^{-1}\, ,
\end{align}
where $\int d{\bf k}$ is a short-hand notation for  $\int d^3 k/(2\pi)^3$.

Eq.~(\ref{Eq_self-consist_DMFT}) can be considered as a self-consistent equation for $\mathbf{\underline{\underline{K}}}^{ab}$ since the direct correlation functions of the single-site effective model are obtained from the action in Eq.~(\ref{Sloc}) as functions of the replica `magnetizations' and of the Weiss fields: $\mathbf{\underline{\underline{C}}}_{\:0}^{ab} \equiv \mathbf{\underline{\underline{C}}}_{\:0}^{ab}[\{m^e, \hat m^e\};\{ \mathbf{\underline{\underline{K}}}^{ef}\}]$. Again, we can use the expansion in number of free replica sums to derive explicit expressions. We decompose the matrices according to $\mathbf{\underline{\underline{K}}}^{ab}=\mathbf{\underline{\underline{K}}}_c^{a}\delta_{ab} + \mathbf{\underline{\underline{K}}}_{\:d}^{ab}$, etc..., and expand each component as in Eqs.~(\ref{Eq_free_replica_Gc},\ref{Eq_free_replica_Gd}) (fixing the magnetizations instead of the sources). At zeroth-order, one finds from Eq.~(\ref{Eq_self-consist_DMFT})
\begin{align}
\label{Eq_self-consist_Cc}
\mathbf{\underline{\underline{G}}}_{\:0,c}(m^a,\hat{m}^a)=\mathbf{\underline{\underline{C}}}_{\:0,c}(m^a,\hat{m}^a)^{-1}= \int d{\bf k} \left[\mathbf{\underline{\underline{C}}}_{\:c}^{(0)}(\mathbf k)+ \mathbf{\underline{\underline K}}_{\:c}(m^a,\hat{m}^a) +\mathbf{\underline{\underline{C}}}_{\:0,c}(m^a,\hat{m}^a)\right] ^{-1}
\end{align}
and
\begin{equation}
\begin{split}
\label{Eq_self-consist_Cd}
- &\mathbf{\underline{\underline{G}}}_{\:0,d}(m^a,\hat{m}^a;m^b,\hat{m}^b)=
\mathbf{\underline{\underline{C}}}_{\:0,c}(m^a,\hat{m}^a)^{-1} \mathbf{\underline{\underline{C}}}_{\:0,d}(m^a,\hat{m}^a; m^b,\hat{m}^b) \mathbf{\underline{\underline{C}}}_{\:0,c}(m^b,\hat{m}^b)^{-1}\\&
= \int d{\bf k}\, \left[\mathbf{\underline{\underline{C}}}_{\:c}^{(0)}(\mathbf k)+ \mathbf{\underline{\underline K}}_{\:c}(m^a,\hat{m}^a) +\mathbf{\underline{\underline{C}}}_{\:0,c}(m^a,\hat{m}^a)\right] ^{-1}
\left[\mathbf{\underline{\underline{C}}}_{\:d}^{(0)}+ \mathbf{\underline{\underline K}}_{\:d}(m^a,\hat{m}^a;m^b,\hat{m}^b) +\mathbf{\underline{\underline{C}}}_{\:0,d}(m^a,\hat{m}^a;m^b,\hat{m}^b)\right]\\&
\times \left[\mathbf{\underline{\underline{C}}}_{\:c}^{(0)}(\mathbf k)+ \mathbf{\underline{\underline K}}_{\:c}(m^b,\hat{m}^b) +\mathbf{\underline{\underline{C}}}_{\:0,c}(m^b,\hat{m}^b)\right] ^{-1} \ ,
\end{split}
\end{equation}
where $\mathbf{\underline{\underline{C}}}_{\:0,c}(m^a,\hat{m}^a)$ and $\mathbf{\underline{\underline{C}}}_{\:0,d}(m^a,\hat{m}^a; m^b,\hat{m}^b)$ are calculated from the single-site effective action in Eq.~(\ref{Sloc}) with the Weiss fields now functions of the magnetizations through the above implicit equations; the latter are  considered at zeroth-order in the expansion in number of free replica sums, \textit{i.e.} $\mathbf{\underline{\underline K}}_{\:c}^a\equiv \mathbf{\underline{\underline K}}_{\:c}(m^a,\hat{m}^a)$ and $\mathbf{\underline{\underline K}}_{\:d}^{ab}\equiv \mathbf{\underline{\underline K}}_{\:d}(m^a,\hat{m}^a;m^b,\hat{m}^b)$.

As the Luttinger-Ward functional is universal, the expression of $\Gamma_{rep}$  is obtained by subtracting from Eq.~(\ref{Eqfunc}) the formal expression of $\Gamma_{rep,0}$ for the single-site effective model. Restricting again the calculation to the translationally invariant situation (uniform configurations), we obtain 
\begin{align}
\label{Eqfunc1}
\frac{1}{N}\Gamma_{rep}&=\Gamma_{rep,0}-qJ\sum_a m^a\hat{m}^a +\frac{1}{2}\sum_{a,b}\left[ K^{ab}m^am^b+\hat{K}^{ab}(m^a\hat{m}^b+\hat{m}^am^b)+(\hat{\hat{K}}^{ab}-\Delta)\hat{m}^a\hat{m}^b\right] \nonumber\\
&-\frac{1}{2}\int d{\bf k} \ \left[ \mbox{Tr}\ln \big({\bf \underline{\underline{G}}}({\bf k}){\bf \underline{\underline{G}}}_{\:0}^{-1}\big)-\mbox{Tr}\,{\bf \underline{\underline{C}}}^{(0)}({\bf k}){\bf \underline{\underline{G}}}({\bf k})+\mbox{Tr}\,{\bf \underline{\underline{C}}}_{\:0}^{(0)}{\bf \underline{\underline{G}}}_{\:0}\right]  .
\end{align}
This effective action, when evaluated at the extremum corresponding to the solution of the Schwinger-Dyson equations, is a function of the replica magnetizations that it can also be expanded in number of free replica sums,
\begin{align}
\label{Eq_Gamma_freereplicasum}
\Gamma_{rep}(\{ m^a, \hat{m}^a\})&=\sum_a \Gamma_{1}(m^a, \hat{m}^a)-\frac{1}{2}\sum_{a,b} \Gamma_{2}(m^a, \hat{m}^a;m^b, \hat{m}^b)+ \cdots \, ,
\end{align}
the $p$th term then providing information on the $p$th cumulant (see above).

Since the local approximation is performed at the level of the Luttinger-Ward functional, the theory possesses some form of thermodynamic consistency\cite{B1962}. Indeed, one obtains the same value of $\Gamma_{rep}$ by using the above equation or by integrating \textit{once} a quantity that depends on the Green's functions (\textit{e.g.} the `internal energy') with respect to the model parameters (\textit{e.g.} the coupling constants). For instance, using the fact that the functional is extremal with respect to the variations of the Green's functions and considering for simplicity the case where all replica magnetizations are equal, one readily derives the relations
\begin{align}
\frac{\partial \Gamma_1/N}{\partial (qJ)}&=-\left[ \hat{G}({\bf r=e})+m\hat{m}\right] \nonumber\\
\frac{\partial \Gamma_1/N}{\partial \Delta}&=-\frac{1}{2}\hat{\hat{G}}_c({\bf r=0}),
\end{align}
which are exact and are thus preserved by the local approximation. The extremal property also allows us to simply relate the fields $H(m,\hat{m})$ and ${\hat H}(m,\hat{m})$ to the fields $H_0(m,\hat{m})$ and ${\hat H}_0(m,\hat{m})$ in the single-site effective system (they all correspond to zeroth-order terms in the expansion in free replica sums) by  only considering the {\it explicit} dependence of $\Gamma$ and $\Gamma_0$ on $m$ and $\hat{m}$. This yields
\begin{align}
H(m,\hat{m})&\equiv\frac{1}{N}\frac{\partial \Gamma_1}{\partial \hat{m}}=\frac{1}{N}\frac{\partial S(m,\hat{m})}{\partial \hat{m}}\nonumber\\
H_0(m,\hat{m})&\equiv \frac{\partial \Gamma_{0,1}}{\partial \hat{m}}\big\vert_{\mathbf{\underline{\underline K}}}=\frac{\partial S_0(m,\hat{m})}{\partial \hat{m}}\, ,
\end{align}
which leads to
\begin{align}
\label{Eqfields1}
H(m,\hat{m})=H_0(m,\hat{m})+[\hat{K}_{c}(m,\hat{m})-qJ]m+\hat{\hat{K}}_{c}(m,\hat{m})  \hat{m} \ .
\end{align}
Similarly, one finds that
\begin{align}
\label{Eqfields2}
\hat{H}(m,\hat{m})=\hat{H}_0(m,\hat{m})+[\hat{K}_{c}(m,\hat{m})-qJ]\hat{m}+K_{c}(m,\hat{m}) m \ .
\end{align}
One may however notice that the local approximation does not really yield a fully consistent theory. For instance, the value of $\hat{C}_c({\bf k=0})$  obtained from Eqs.~(\ref{Eq_SD_fourier},\ref{Eq_schwinger-dyson_0},\ref{Eq_self-consist_DMFT}) does not coincide with that  obtained from the `susceptibility sum-rule' $\hat{C}_c({\bf k=0})=\partial^2 (\Gamma_1/N)/\partial m\partial \hat{m}=\partial H(m,\hat{m})/\partial m$. This inconsistency is a well-known flaw of this type of approximation  and it can be cured by renormalizing the value of the direct correlation functions at nearest-neighbor distance as is done for instance in the so-called Self-Consistent Ornstein-Zernike Approximation (SCOZA)\cite{KRT1999,HS1977}. This route, however, looks prohibitively difficult in the present case and will not be pursued. Improving the theory by introducing a ${\bf k}$-dependence in the self-energies as is done in the cluster dynamical mean-field theory\cite{GKKR1996} can make the theory exact up to order $1/d$, which is an interesting property, but it does not solve the above inconsistency problem.

At this stage, we must point out a serious difficulty occuring in the approximate framework developed above. We have stressed that, just like the reference system, the exact system is most likely such that the Green's function $\hat{\hat{G}}_d$ is singular with a diverging term proportional to $\delta(0)$. The 2PI functional and the other Green's functions being most likely finite, the contribution of $\hat{\hat{G}}_d$ must altogether vanish in the expressions of these finite quantities. One can therefore plainly drop all dependence on $\hat{\hat{G}}_d$ in Eq.~(\ref{Eqfunc}), which amounts to a perfect cancellation between terms in $\mbox{Tr}\ln {\bf \underline{\underline{G}}}^{-1}$ and in $\Phi_{rep}$ (there are no $\hat{\hat{G}}_d$ contributions in $\mbox{Tr}\,{\bf \underline{\underline{C}}}^{(0)}{\bf \underline{\underline{G}}}$). This of course has consequences on the self-energies. To avoid inconsistencies, this property must be satisfied, or at least enforced, in any sensible approximation. As shown in the appendix, it is easily realized that $\hat{\hat{G}}_{0,d}$ in the single-site model may develop a singular $\delta(0)$ term only if the Weiss field $\hat{\hat{K}}_{c}$ is identically zero. A sensible approximation scheme must therefore be compatible with (i) setting  $\hat{\hat{K}}_{c}$ to zero, (ii) discarding the self-consistent equation on $\hat{\hat{G}}_d$, and (iii) dropping all contributions involving $\hat{\hat{G}}_d$ in the two-particle irreducible functional of the original model and of the effective single-site model. It turns out that these requirements are not met by the local self-energy approximation inside the hysteresis loop (when $\hat{H}$ and $\hat m$ are finite and the quenched complexity is strictly positive). Along the hysteresis loop  ($\hat{H},\hat m \rightarrow \pm \infty$), the situation improves and the local self-energy approximation is well behaved, at least at the level of single-replica quantities. Awaiting for a proper resolution in the general case (see the discussion in conclusion), we now discuss the behavior on the hysteresis loop.

\subsection{Behavior along the hysteresis loop}

We consider the solution of the LSEA discussed above along the hysteresis loop, when $\hat{H}\rightarrow  \pm \infty$ or alternatively when $\hat{m}\rightarrow  \pm \infty$. Our main assumptions, motivated by the expected exact behavior along the hysteresis loop, are that the  Weiss field $\hat{\hat{K}}_{c}$ and the corresponding direct correlation function $\hat{\hat{C}}_{0,c}$ in the effective model are  identically zero (see above) and that the complexity is zero as well. The former of these two assumptions implies from Eq. (\ref{EqOZc}) that $\hat{\hat{C}}_c({\bf k})$ and thus $G_c({\bf k})$ in the original model are also zero, as anticipated. We are then only interested in the correlation functions that are physically observable, namely $\hat{G}_c({\bf k})$ and $G_d({\bf k})$. As a result, we only need to solve the two corresponding self-consistency equations, $\hat{G}_{c}(\mathbf r = \mathbf 0; m^a)=\hat{G}_{0,c}(m^a)$ and $G_{d}(\mathbf r = \mathbf 0; m^a;m^b)=G_{0,d}(m^a;m^b)$, leading to
\begin{align}
\label{simplifiedOZ_G0c}
\hat{G}_{0,c}(m^a)=\int d{\bf k}  \frac{1}{\hat{C}_{0,c}(m^a)+\hat{K}_{c}(m^a)-qJ\lambda({\bf k})}
\end{align}
\begin{align}
\label{simplifiedOZ_G0d}
G_{0,d}(m^a;m^b)=-\int d{\bf k}  \frac{\hat{\hat{C}}_{0,d}(m^a;m^b)+\hat{\hat{K}}_{d}(m^a;m^b)-\Delta}{[\hat{C}_{0,c}(m^a)+\hat{K}_{c}(m^a)-qJ\lambda({\bf k})][\hat{C}_{0,c}(m^b)+\hat{K}_{c}(m^b)-qJ\lambda({\bf k})]}
\end{align}
where $\hat{G}_{0,c}(m)$, $\hat{C}_{0,c}(m)=\hat{G}_{0,c}^{-1}(m)$, $G_{0,d}(m^a,m^b)$, and $\hat{\hat{C}}_{0,d}(m^a;m^b)=-\hat{C}_{0,c}(m^a)G_{0,d}(m^a,m^b)\hat{C}_{0,c}(m^b)$ are obtained from the effective single-site action
\begin{align}
\label{Sloc_Zeroth_order}
S_{0,rep}[\{s^a,\hat{s}^a\}]= \sum_{a}\left[ - \hat{K}_c^{a}s^a + V'(s^{a})\right] \hat{s}^a - \frac{1}{2}\sum_{a,b}\hat{\hat{K}}_d^{ab} \hat{s}^a\hat{s}^b,
\end{align}
with the magnetizations $m^a$ fixed by the external sources $H_0(m^a)$ [$\hat{m}^a\rightarrow \pm \infty$ follows from $\hat{H}_0(m^a)\rightarrow \pm \infty$] and the Weiss fields implicitly determined as functions of the magnetizations, $\hat{K}_c^{a}\equiv\hat{K}_c(m^{a})$, $\hat{\hat{K}}_d^{ab}\equiv \hat{\hat{K}}_d(m^a;m^b)$. The actual computation of the correlation functions associated to this effective single-site action is performed in the appendix.
After introducing
\begin{align}
z(m)=\frac{qJ}{\hat{C}_{0,c}(m)+\hat{K}_{c}(m)},
\end{align}
$P(z)\equiv P({\bf r=0};z)$, and $P'(z)=dP(z)/dz$ (as in section IV), Eqs.~(\ref{simplifiedOZ_G0c}) and (\ref{simplifiedOZ_G0d}) can be rewritten as
\begin{align}
\label{hGc}
\frac{1}{\hat{C}_{0,c}(m)}=\frac{1}{\hat{C}_{0,c}(m)+\hat{K}_{c}(m)}P(z(m)),
\end{align}
where we have dropped the superscript $a$ on the magnetization, and
\begin{align}
\label{Gd_2copies}
\frac{\hat{\hat{C}}_{0,d}(m^a;m^b)}{\hat{C}_{0,c}(m^a)\hat{C}_{0,c}(m^b)}=\frac{\hat{\hat{C}}_{0,d}(m^a;m^b)+\hat{\hat{K}}_{d}(m^a;m^b)-\Delta}{[\hat{C}_{0,c}(m^a)+\hat{K}_{c}(m^a)][\hat{C}_{0,c}(m^b)+\hat{K}_{c}(m^b)]}\, \left[ \frac{z(m^a)P(z(m^a))- z(m^b)P(z(m^b))}{z(m^a)-z(m^b)}\right] .
\end{align}
For equal magnetizations $m^a=m^b=m$, Eq.~(\ref{Gd_2copies}) simplifies to
\begin{align}
\label{Gd}
\frac{\hat{\hat{C}}_{0,d}(m;m)}{\hat{C}_{0,c}(m)^2}=\frac{\hat{\hat{C}}_{0,d}(m;m)+\hat{\hat{K}}_{d}(m;m)-\Delta}{[\hat{C}_{0,c}(m)+\hat{K}_{c}(m)]^2}\left[ P(z(m))+z(m)P'(z(m))\right] .
\end{align}

Once the coupled equations (\ref{hGc}) and (\ref{Gd}) are solved for $\hat{K}_{c}(m)$ and $\hat{\hat{K}}_{d}(m;m)$, we calculate the field $H$ along the loop from Eq.~(\ref{Eqfields1}) with $\hat{\hat{K}}_{c}=0$, 
\begin{align}
\label{Eq_H_hysteresis}
H(m)=H_0(m)+[\hat{K}_{c}(m)-qJ]m, 
\end{align}
and  we obtain the physical correlation (Green's) functions for equal magnetizations from Eqs. (\ref{RPA1}) as
\begin{align}
\label{Gsh_1copy}
G_{sh}(\mathbf{r};m)=\Delta\; \hat{G}_{0,c}(m)\frac{P(\mathbf{r};z(m))}{P(z(m))}
\end{align}
\begin{align}
\label{Gss_1copy}
G_{ss}(\mathbf{r};m)=G_{0,d}(m;m)\frac{P(\mathbf{r};z(m))+z(m) P'(\mathbf{r};z(m))}{P(z(m))+z(m)P'(z(m))} \ , 
\end{align}
with explicit expressions for $\hat{G}_{0,c}(m)$ and $G_{0,d}(m;m)$ given in the appendix .

\begin{figure}[hbt]
\begin{center}
\includegraphics[width=9cm]{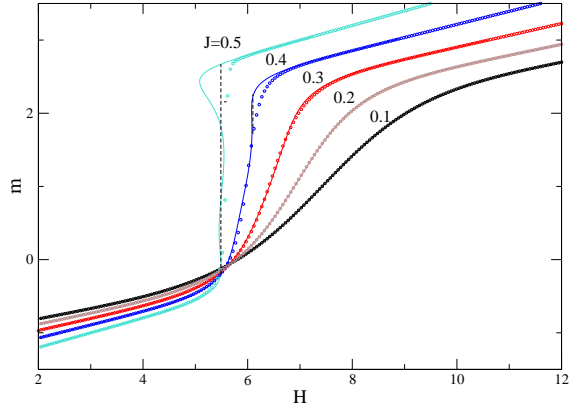}
\caption{\label{Fig7} Magnetization along the ascending branch of the hysteresis loop: comparison of the predictions of the local self-energy approximation (solid lines)  to the simulation data (circles) for various values of the coupling $J$. The dashed lines for $J=0.4$ and $J=0.5$ indicate the macroscopic jump predicted by the theory.}
\end{center}
\end{figure}

\begin{figure}[hbt]
\begin{center}
\includegraphics[width=7.5cm]{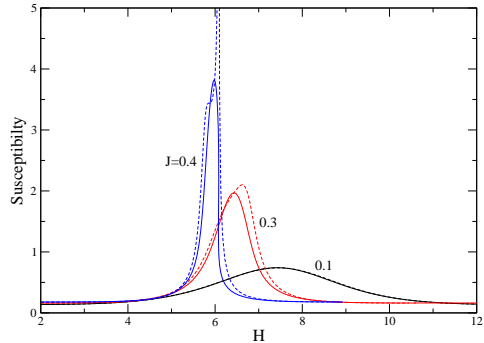}
\caption{\label{Fig8} Test of thermodynamic self-consistency in the LSEA: comparison of the susceptibility $\hat{G}_c({\bf k=0})$  (solid lines) with the slope  $dm/dH$ of the magnetization curve (dashed lines) along the ascending branch of the hysteresis curve. For $J=0.4$ the slope diverges at the spinodal field $H_{sp}\approx 6.08$ whereas $\hat{G}_c({\bf k=0})$ remains finite.}
\end{center}
\end{figure}

The predictions of Eq. (\ref{Eq_H_hysteresis}) for the ascending branch of the hysteresis loop are shown in Fig. \ref{Fig7}. Comparing with Fig. \ref{Fig2}, we see that the improvement over the RPA is quite significant: the agreement with the simulations is now satisfactory up to $J=0.4$ where a small reentrant behavior is observed in the upper part of the curve. This behavior, which erroneously indicates that the system has entered the small-disorder regime, is related to the fact that the effective action is obtained via the `energy' route which is not consistent with the `susceptibility' one, as mentioned above. This lack to thermodynamic consistency as the coupling increases is illustrated in Fig. \ref{Fig8}. Of course, the actual magnetization cannot decrease as $H$ increases and it must jump at a `spinodal'  field where the slope $dm/dh$ diverges for the first time (as can be seen in Fig.~\ref{Fig7} for $J=0.5$, the theoretical curve may display several spinodal fields). Note that due to this inconsistency the condition $dm/dH\rightarrow \infty$ does not imply $\hat{G}_c({\bf k=0})=0$ and thus not $z=1$ (from Eq.~(\ref{Gsh_1copy}) one has $\hat{G}_c({\bf k=0})=(1/\Delta) G_{sh}({\bf k=0})=\hat{G}_{0,c}/[P(z)(1-z)]$).

\begin{figure}[hbt]
\begin{center}
\includegraphics[width=9cm]{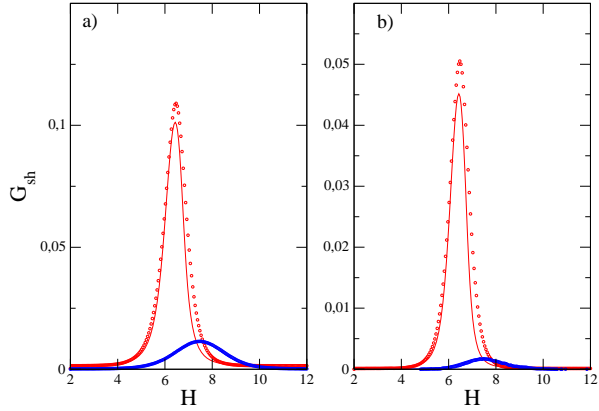}
\caption{\label{Fig9} Spin-random field correlation function at the second (a) and third (b) nearest-neighbors along the ascending branch of the hysteresis curve for $J=0.1$ (blue) and $J=0.3$ (red). The predictions of the LSEA (solid lines) are compared to the simulation data (circles).}
\end{center}
\end{figure}

\begin{figure}[hbt]
\begin{center}
\includegraphics[width=9cm]{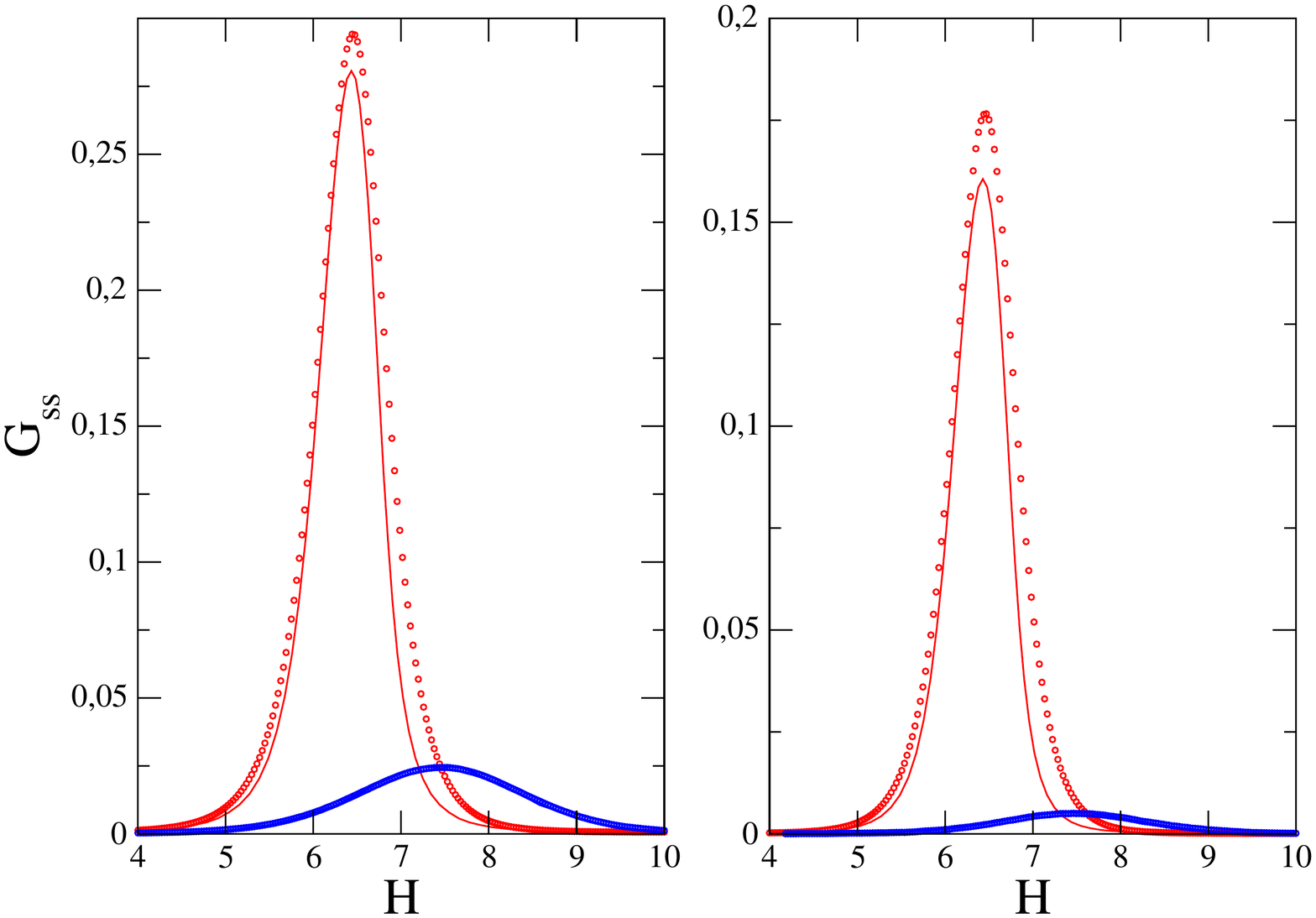}
\caption{\label{Fig10} Same as Fig. \ref{Fig9}. for the spin-spin correlation function $G_{ss}({\bf r})$.}
\end{center}
\end{figure}

The correlation functions $G_{sh}({\bf r})$ and $G_{ss}({\bf r})$ at ${\bf r=0}$ and ${\bf r=e}$  are compared to the simulation data and to  the RPA in Figs. \ref{Fig3} and \ref{Fig4}. The agreement with the simulations is now excellent for $J=0.3$.  In particular, the LSEA correctly predicts that the maximum of $G_{sh}(\textbf{r=0})$ decreases as $J$ increases, contrary to the RPA. The values at the second and third nearest-neighbor distances\cite{note4} are also shown in Figs. \ref{Fig9} and \ref{Fig10}. The theoretical predictions for $J=0.3$ are still fairly good although the values of the functions are slightly overestimated. More generally, the results displayed in these figures show that the actual dependence of $G_{sh}({\bf r}) $ and $G_{ss}({\bf r})$ with distance is well described by RPA-like expressions in the large-disorder regime and not too close to criticality. The two functions   involve a single correlation length $\xi$ [{\it e.g.} the second-moment correlation length defined by $G_{sh}({\bf k}) \sim G_{sh}(0)(1 + \xi^2k^2)$, $k \rightarrow 0$, and thus related to $z$ by $q\xi^2 = z/(1-z)$] and the main effect of disorder fluctuations is to renormalize $\xi$.

\begin{figure}[hbt]
\begin{center}
\includegraphics[width=9cm]{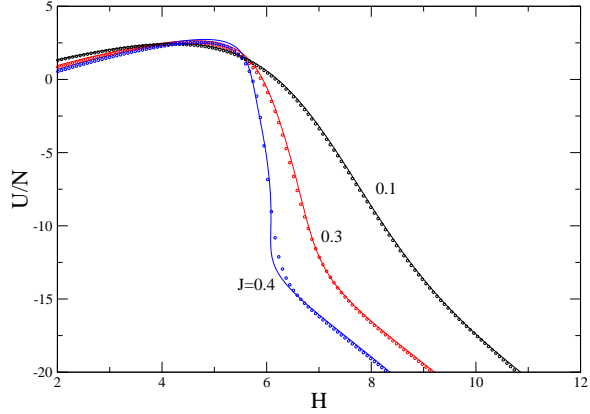}
\caption{\label{Fig11} Energy per spin along the ascending branch of the hysteresis loop. The predictions of the LSEA (solid lines) are compared to the simulation data (circles). }
\end{center}
\end{figure}

From the above correlation functions, we can compute the average energy per spin along the loop, which is the sum of four contributions,
\begin{align}
U/N=\overline{{\cal H}}/N=u_{1}+u_{2}+u_{3}+u_{4}
\end{align}
with
\begin{align}
u_{1}&=-\frac{J}{N}\sum_{<i,j>}\overline{s_is_j}=-\frac{qJ}{2}[G_{ss}(r=1)+m^2]\nonumber\\
u_2&=-\overline{s_i}H=-mH\nonumber\\
u_3&=-\overline{s_ih_i}=-G_{sh}(r=0)\nonumber\\
u_4&=\overline{V(s_i)}=\frac{1}{2k}\overline{(H+J\sum_{j/i}s_j+h_i)^2}=\frac{1}{2k}\big[H^2+\Delta+2qJmH+2qJG_{sh}(r=1)\nonumber\\
&+qJ^2(G_{ss}(r=0)+m^2)+q(q-2)J^2(G_{ss}(r=\sqrt{2})+m^2)+qJ^2(G_{ss}(r=2)+m^2)\big] \ ,
\end{align}
where, again, the dependence of $H$ and of the Green's functions on the magnetization $m$ is left implicit. As can be seen in Fig.~\ref{Fig11}, the average energy per spin is also very well reproduced for $J=0.3$ and the discrepancies for $J=0.4$ are limited to the small range of $H$ where the reentrant behavior in the magnetization occurs.

\subsection{Two-replica correlation function and avalanches along the hysteresis loop}

Additional information on the system along the hysteresis loop is encoded in the $2$-replica spin-spin correlation function for distinct magnetizations, $G_{ss}(\mathbf{r};m^a, m^b)\equiv \lim_{\hat{H}\rightarrow \pm \infty}G_{d}(\mathbf{r};m^a, m^b))=\overline {s_0(H^a)s_{\bf r}(H^b)}- m(H^a)m(H^b)$. From the equations above, one derives
\begin{align}
\label{Gss_2copies}
G_{ss}(\mathbf{r};m^a;m^b)= G_{0,d}(m^a;m^b)\left[ \frac{z(m^a)P(\mathbf{r};z(m^a))- z(m^b)P(\mathbf{r};z(m^b))}{z(m^a)P(z(m^a))-z(m^b)P(z(m^b))}\right] ,
\end{align}
which gives back Eq.~(\ref{Gss_1copy}) when the magnetizations are equal.
As discussed in section III, this function allows one to compute the unnormalized second moment of the avalanche-size distribution  via Eq. (\ref{S2moment}), provided one extracts the nonanalytic cusp-like dependence of $G_{ss}(\mathbf{k}=\mathbf{0},m^a;m^b)$ on the difference $m^a-m^b$, and consequently on $H^a -H^b$. The amplitude of the linear-cusp contribution is then obtained from $\partial G_{ss}(\mathbf{k}=\mathbf{0};m^a;m^b))/\partial(m^a-m^b)$ when $(m^a-m^b) \rightarrow 0^+$. It is clear from Eq.~(\ref{Gss_2copies}) that the cusp only comes from $G_{0,d}(m^a;m^b)$, \textit{i.e.}  from the two-replica correlation function in the  single-site effective model defined by the action in Eq.~(\ref{Sloc_Zeroth_order}). This function depends on the magnetizations both in an explicit way and through the Weiss fields, which we write as $G_{0,d}(m^a;m^b)\equiv G_{0,d}(m^a;m^b; \hat K_c(m^a),\hat K_c(m^b),\hat{\hat K}_d(m^a;m^a),\hat{\hat K}_d(m^b;m^b),\hat{\hat K}_d(m^a;m^b))$. The contribution of the Weiss fields to the cusp can only come from $\hat{\hat K}_d(m^a;m^b)$ [with $m^a\neq m^b$], implying  that 
\begin{align}
\label{Gd0_cusp}
\frac{\partial G_{0,d}}{\partial(m^a-m^b)}\bigg|_{0^+}=\frac{\partial G_{0,d}}{\partial(m^a-m^b)}\bigg|_{\hat K_c,\hat{\hat K}_d;0^+}+\frac{\partial G_{0,d}}{\partial \hat{\hat K}_d(m^a;m^b)}\bigg|_{0^+} \frac{\partial \hat{\hat K}_d(m^a;m^b)}{\partial(m^a-m^b)}\bigg|_{0^+} \ ,
\end{align}
where the subscript $0^+$ indicates that the derivatives are evaluated for $(m^a-m^b) \rightarrow 0^+$ and the derivative with respect to $\hat{\hat K}_d(m^a;m^b)$  in the right-hand side is taken with $m^a,m^b$, $\hat K_c(m^a),\hat K_c(m^b)$, and $\hat{\hat K}_d(m^a;m^a),\hat{\hat K}_d(m^b;m^b)$ fixed. 
On the other hand, the self-consistency equation (\ref{simplifiedOZ_G0d}) imposes that the cusp in  $G_{0,d}(m^a;m^b)$ is directly proportional to the cusp in the Weiss field  $\hat{\hat K}_d(m^a;m^b)$. Indeed, replacing $\hat{\hat{C}}_{0,d}(m^a;m^b)$ by $-\hat{C}_{0,c}(m^a)G_{0,d}(m^a;m^b)\hat{C}_{0,c}(m^b)$ in the right-hand side of Eq. (\ref{simplifiedOZ_G0d}),  one finds after simple manipulations 
\begin{align}
\label{hathatKd0_cusp}
\frac{\partial G_{0,d}}{\partial(m^a-m^b)}\bigg|_{0^+}= \left[\frac{ G_{0,d}(m,m)}{\hat{\hat K}_d(m;m)- \Delta}\right] \frac{\partial \hat{\hat K}_d(m^a;m^b)}{\partial(m^a-m^b)}\bigg|_{0^+} 
\end{align}
where $m$ is the common limit of $m^a$ and $m^b$.

Before we proceed any further, we must however deal with a difficulty that arises in Eq.~(\ref{Gd0_cusp}) from the behavior of the partial derivative $\partial G_{0,d}/ \partial \hat{\hat K}_d(m^a;m^b)$ when $(m^a-m^b) \rightarrow 0^+$. At zeroth-order in the expansion in number of free replica sums, it is equivalent to use the replica magnetizations $(\{m^a\},\{m^a\}$) or the corresponding sources $(\{H_0^a=H_0(m^a)\},\{\hat{H}_0^a=\hat{H}_0(m^a)\}$), so that we may  as well consider the partial derivative of $G_{0,d}(H_0^a,H_0^b)$ with respect to $\hat{\hat K}_d(H_0^a;H_0^b)$. This quantity is most easily expressed by using the property that $G_{0,d}(H_0^a,\hat{H}_0^a;H_0^b,\hat{H}_0^b)$ is the second derivative of the second cumulant $W_2(H_0^a,\hat{H}_0^a;H_0^b,\hat{H}_0^b)$ with respect to $\hat {H}_0^a$ and $\hat {H}_0^b$. On the other hand, from the definition of the action of the effective model, the derivative of $W_2$ with respect to $\hat{\hat{K}}_{d}^{ab}=\hat{\hat{K}}_{d}(H_0^a,H_0^b)$  is equal to $\hat{\hat{G}}_{0,d}(H_0^a,\hat H_0^a;H_0^b,\hat H_0^b)+\hat m(H_0^a,\hat H_0^a) \hat m(H_0^b,\hat H_0^b)$ (using the symmetry $\hat{\hat{K}}_{d}^{ab}=\hat{\hat{K}}_{d}^{ba}$). After permuting the order of the derivatives, one thus has
\begin{align}
\label{derivGK}
\frac{\partial G_{0,d}}{\partial \hat{\hat K}_d(H_0^a;H_0^b)}=\frac{\partial^2}{\partial \hat{H}_0^a\partial \hat{H}_0^b}[\hat{\hat{G}}_{0,d}(H_0^a,\hat H_0^a;H_0^b,\hat H_0^b)+\hat m(H_0^a,\hat H_0^a) \hat m(H_0^b,\hat H_0^b)] \ ,
\end{align}
a quantity that is still defined in the limit $\hat{H}_0^a,\hat{H}_0^b\rightarrow \pm \infty$ since $\hat{\hat{G}}_{0,d}(H_0^a,\hat H_0^a;H_0^b,\hat H_0^b)+\hat m(H_0^a,\hat H_0^a) \hat m(H_0^b,\hat H_0^b)$ is proportional to $\hat{H}_0^a\hat{H}_0^b$   [see Eqs.~(\ref{EqGd0reg}) and (\ref{EqGd0sing})]. The problem is that the Green function $\hat{\hat{G}}_{0,d}$ contains a singular term proportional to $\delta(H_0^a-H_0^b)$, as already mentioned, so that $\partial G_{0,d}/\partial \hat{\hat K}_d(H_0^a;H_0^b)$ diverges when $H_0^a-H_0^b\rightarrow 0$ or, equivalently, when $m^a-m^b\rightarrow 0$. This causes the right-hand side of Eq.~(\ref{Gd0_cusp}) to diverge, whereas Eq.~(\ref{hathatKd0_cusp}) does not predict any divergence. This spurious behavior is intrinsic to the local self-energy approximation but it can be circumvented by simply discarding the singular contribution of $\hat{\hat{G}}_{0,d}$ and only keeping in Eq.~(\ref{derivGK}) the contribution of the regular term. This is admittedly a drastic regularization procedure which however may be rationalized by recalling that the singular term, which is an exact feature of the Green function $\hat{\hat{G}}_{d}$, should not play any role in an exact treatment\cite{note6}. Then, combining Eqs.~(\ref{Gd0_cusp}) and (\ref{hathatKd0_cusp}) and considering the fields instead of the magnetizations lead to
\begin{align}
\label{Gd0_cusp_final}
\frac{\partial G_{0,d}(H_0^a;H_0^b)}{\partial(H_0^a-H_0^b)}\bigg|_{0^+}=\left(1-  \left[ \frac{\hat{\hat K}_d(H_0;H_0)- \Delta}{G_{0,d}(H_0,H_0)}\right] \frac{\partial G_{0,d}}{\partial \hat{\hat K}_d(H_0^a;H_0^b)}\bigg|_{0^+}^{reg}\right)^{-1} \frac{\partial G_{0,d}}{\partial(H_0^a-H_0^b)}\bigg|_{\hat K_c,\hat{\hat K}_d;0^+} \ . 
\end{align}
where $H_0=H_0(m)$ is the common limit of $H_0^a$ and $H_0^b$. It is important to note that the two derivatives of $G_{0,d}$ that appear in the right-hand side can now be computed from the effective single-site action in Eq.~(\ref{Sloc_Zeroth_order}) by  assuming that the Weiss fields are replica-symmetric and independent of the sources $H_0$ and $\hat{H}_0$. This calculation is performed in the appendix.

Once the amplitude of the cusp in $|m^a-m^b|$ is known, the amplitude of the cusp in $|H^a-H^b|$, $G_{ss}^{cusp}({\bf k=0},H)$, which is needed for computing the unnormalized second moment of the avalanche distribution via Eq.~(\ref{S2moment}), is obtained through
\begin{align}
\label{Gd_cuspH}
G_{ss}^{cusp}({\bf k=0},H)=\frac{\partial G_{ss}({\bf k=0};m^a;m^b)}{\partial(m^a-m^b)}\bigg|_{0^+} \left( \frac{\partial m}{\partial H}\right) ,
\end{align}
with $H$ and $m$ related via Eq.~(\ref{Eq_H_hysteresis}). One finally obtains from the Fourier transform of Eq.~(\ref{Gss_1copy})
\begin{align}
\label{Gd_cuspH_final}
G_{ss}^{cusp}({\bf k=0},H)&=\frac{\partial G_{0,d}(m^a;m^b)}{\partial(m^a-m^b)}\bigg|_{0^+} \frac{(\partial m/\partial H)}{[1-z(m)]^2[P(z(m))+z(m)P'(z(m))]}\nonumber\\
&=\frac{\partial G_{0,d}(H_0^a;H_0^b)}{\partial(H_0^a-H_0^b)}\bigg|_{0^+} \frac{(\partial m/\partial H)/(\partial m/\partial H_0)}{[1-z(m)]^2[P(z(m))+z(m)P'(z(m))]} \ ,
\end{align}
with $\partial G_{0,d}(H_0^a;H_0^b)/\partial(H_0^a-H_0^b)\big|_{0^+}$ given by Eq.~(\ref{Gd0_cusp_final}). This equation further simplifies if one replaces $(\partial m/\partial H)$ by $\hat G_c({\bf k=0})=\hat G_{0,c}/[(1-z)P(z)]$, thereby neglecting the small thermodynamic inconsistency of the present approach. (On the other hand, $(\partial m/\partial H_0)$ is exactly equal to  $\hat G_{0,c}$ in the single-site effective model.)
\begin{figure}[hbt]
\begin{center}
\includegraphics[width=9cm]{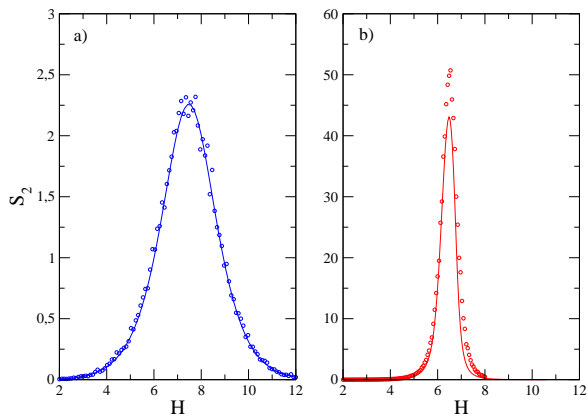}
\caption{\label{Fig12} The unnormalized second moment of the avalanche-size distribution along the ascending branch of the hysteresis curve for $J=0.1$ (a) and $J=0.3$ (b) . The predictions of the LSEA (solid lines) are compared to the simulation data (circles) resulting from an average over $500$ disorder realizations of linear size $L=10$ for $J=0.1$ and $L=20$ for $L=0.3$\: \cite{note5}. }
\end{center}
\end{figure}

The theoretical prediction for the unnormalized second moment of the avalanche-size distribution is compared  to the simulation data in Fig.~\ref{Fig12}.  One observes a reasonable agreement, even for $J=0.3$, which gives some {\it a posteriori} justification to the  regularization procedure that has been used in the above calculation.

\section{Conclusion and outlook}

In this paper we have proposed a formalism to study the out-of-equilibrium hysteresis behavior of random field systems when submitted to an adiabatically varying external field at zero temperature. The key ingredients consist in relating the out-of-equilibrium behavior to the statistics of the metastable states, introducing auxiliary variables to handle the latter, and building an approximation scheme based on the structure of the pair correlation functions. We have applied this program to a soft-spin version of the random field Ising model and focused on describing the system along the hysteresis loop, which is the envelope of the metastable states in the magnetization-applied field plane. The physics of the problem involves `avalanches' between metastable states that in turn generates nonanalyticities in the dependence of several correlation functions on their arguments. We have used an approximation borrowed from condensed-matter theory and most conveniently formulated in a 2PI framework, whose lowest order amounts to neglecting the spatial dependence of the self-energies. We have derived in this way predictions for the physical, spin-spin and spin-random-field, pair correlation functions, as well as for the second moment of the avalanches, which can be compared to computer simulation data: we find a good agreement between predictions and simulation data above the (out-of-equilibrium) critical point, which represents a significant improvement over the mean-field (random-phase approximation) results also calculated here. Away from criticality, the correlation functions along the hysteresis loop keep essentially the same spatial structure as in the random-phase approximation with however a strong renormalization of the correlation length due to disorder-induced fluctuations. 

In spite of the encouraging accuracy of the results, we have encountered difficulties in extending the basic local self-energy approximation either to the situation `inside' the hysteresis loop, for which the complexity associated with the number of metastable states is nonzero, or to the many-replica correlation functions, even along the hysteresis loop. The presence of avalanches at $T=0$ indeed generates a strong singularity in the form of a Dirac delta function in the (unphysical) correlation function of the auxiliary variable. The contribution of the latter must therefore exactly vanish in physical quantities and in the effective action. Such exact cancellations are however hard to implement in an approximate theory. To cure this problem, one must somewhat relax the self-consistency of the local self-energy approximation to allow for one more constraint enforcing the vanishing of the diverging terms without overconstraining the theory.  In addition, to improve the accuracy of the predictions and, for instance, to provide a good description of the much studied hard-spin Ising model in a random field, one will have to go beyond the local approximation for the self-energies, as done in the cluster dynamical mean-field theory\cite{GKKR1996}. Work in this direction is in progress. The vicinity of the out-of-equilibrium critical point on the other hand cannot be treated in such cluster extensions of the local approximation and requires a different treatment along the lines of the recently developed nonperturbative functional renormalization group\cite{TT2008}.

\appendix 

\section{Single-site effective model with `replica-symmetric' Weiss fields}

In this appendix we compute the correlation (Green's) functions of the single-site effective model whose partition function in replica space reads
\begin{equation}
\label{B1}
{\cal Z}_{rep}(\{H^a,\hat{H}^a\})= \int \prod_a d s^a d\hat{s}^a\ e^{-S_{rep}[\{s^a,\hat{s}^a\}]+\sum_a[\hat{H}^a s^a+H^a\hat{s}^a]} 
\end{equation}
where the action is given Eq. (\ref{Sloc}) (hereafter, for ease of notation we drop the subscript $0$ on all quantities.) We consider a replica-symmetry ansatz for the Weiss fields, $\underline{\underline{K}}^{ab}=\underline{\underline{K}}_{c}\delta_{ab}+\underline{\underline{K}}_{d}$ with $\underline{\underline{K}}_{c}$ and $\underline{\underline{K}}_{d}$ taken as fixed.  Although this ansatz neglects the fact that the Weiss fields depend on the magnetizations $\{m^a\}$ and $\{\hat{m}^a\}$ through the self-consistency equations (\ref{Eq_self-consist_Cc})-(\ref{Eq_self-consist_Cd}), the results derived in this appendix are nonetheless sufficient to compute all the quantities studied in the present work, including the amplitude of the linear cusp in the two-replica spin-spin correlation function. The action then becomes 
\begin{align}
\label{B2}
S_{rep}(\{s^a,\hat{s}^a\})= -\frac{1}{2}\sum_{a}\big[K_{c} (s^a)^2+2\hat{K}_{c} s^a \hat{s}^a+\hat{\hat{K}}_{c}(\hat{s}^a)^2] -\frac{1}{2}\big[K_{d} u^2+2\hat{K}_{d} uv+\hat{\hat{K}}_{d} v^2\big]+\sum_{a} \hat{s}^{a}V'(s^{a})
\end{align}
where $u=\sum_a s^a$ and $v=\sum_a \hat{s}^a$. The quadratic dependence on $u$ and $v$ can be eliminated by using a Hubbard-Stratonovich transformation with two auxililary fields $h_1$ and $h_2$ that play the role of correlated random fields. This yields
\begin{align}
\label{B3}
{\cal Z}_{rep}(\{H^a, \hat{H}^a\})= \frac{1}{2\pi \sqrt{R_d}}\int dh_1 dh_2\: e^{-\frac{1}{2R_d}\big[\hat{\hat{K}}_{d} h_1^2-2\hat{K}_{d} h_1h_2+K_{d} h_2^2\big]}\prod_a \int d s^a d\hat{s}^a \: e^{f_a(s^a,\hat{s}^a;h_1,h_2)}
\end{align}
where
\begin{align}
\label{B4}
f_a(s^a,\hat{s}^a;h_1,h_2)= \frac{1}{2}\big[K_{c} (s^a)^2+2\hat{K}_{c} s^a \hat{s}^a+\hat{\hat{K}}_{c}(\hat{s}^a)^2] +s^a(\hat{H}_a+h_1)+\hat{s}^{a}[H_a+h_2-V'(s^{a})] 
\end{align}
and $R_d=K_{d} \hat{\hat{K}}_{d}-\hat{K}_{d}^2$ must be a positive quantity.

When all sources act identically in each replica ({\it i.e.} $H^a=H,\hat{H}^a=\hat{H}$), we then have 
\begin{align}
\label{B5}
W_1(H,\hat{H})=\lim_{n\rightarrow 0}\frac{1}{n} \ln {\cal Z}_{rep}(H,\hat{H})= \frac{1}{2\pi \sqrt{R_d}}\int dh_1 dh_2\: e^{-\frac{1}{2R_d}\big[\hat{\hat{K}}_{d} h_1^2-2\hat{K}_{d} h_1h_2+K_{d} h_2^2\big]}{\cal W}(H,\hat{H};h_1,h_2)
\end{align}
with 
\begin{align}
\label{B6}
{\cal W}(H,\hat{H};h_1,h_2)=\ln \int ds \: d\hat{s} \: e^{f(s,\hat{s};h_1,h_2)} \ .
\end{align}

To compute  this quantity we first integrate $e^{f(s,\hat{s})}$ over $\hat{s}$ along the imaginary axis (taking into account the factor $1/(2i\pi)$ that was adsorbed in $d\hat{s}$). This gives
\begin{align}
\int d\hat{s} \: e^{f(s,\hat{s};h_1,h_2)}&=\frac{1}{\sqrt{2\pi\hat{\hat{K}}_{c}}}\exp\big(\frac{g(s;h_1,h_2)}{2\hat{\hat{K}}_{c}} \big)
\end{align}
with
\begin{align}
g(s;h_1,h_2)&=K_{c} \hat{\hat{K}}_{c} s^2 +2\hat{\hat{K}}_{c}(\hat{H}+h_1)s -[H+h_2+\hat{K}_{c} s-V'(s)]^2\nonumber\\
&= [K_{c}\hat{\hat{K}}_{c}-(k-\hat{K}_{c})^2] s^2 +2\big[(\hat{H}+h_1)\hat{\hat{K}}_{c}+(k-\hat{K}_{c})(H+h_2+k\ \mbox{sign}(s))\big]s -(H+h_2+k\ \mbox{sign}(s))^2 \ .
\end{align}
 The second integration over $s$ from $-\infty$ to $0$ and from $0$ to $+\infty$  finally yields 
\begin{align}
\label{EqA9}
{\cal W}(H,\hat{H};h_1,h_2)&= -\frac{1}{2}\ln(R_c)+\ln\big(\frac{1-\mbox{erf}(Y_-)}{2}\ e^{X_-}+\frac{1+\mbox{erf}(Y_+)}{2}\ e^{X_+}\big)
\end{align}
with 
\begin{align}
\label{EqA10}
X_{\pm}= \frac{\hat{\hat{K}}_{c}}{2R_c}(\hat{H}+h_1)^2+\frac{1}{2R_c}(H+h_2\pm k)[K_{c}(H+h_2\pm k)+2(k-\hat{K}_{c})(\hat{H}+h_1)]
\end{align}
and
\begin{align}
\label{EqA11}
Y_{\pm}=\frac{1}{\sqrt{2\hat{\hat{K}}_{c}R_c}} [\hat{\hat{K}}_{c}(\hat{H}+h_1)+(k-\hat{K}_{c})(H+h_2 \pm k)] \ .
\end{align}
Like $R_d$, the quantity $R_c=(k-\hat{K}_{c})^2-K_{c}\hat{\hat{K}}_{c}$ must be positive.  The Green's functions can be calculated by derivation of $W_1(H,\hat{H})$ with respect to the Weiss fields, 
\begin{align}
\label{B7}
\frac{\partial W_1}{\partial K_{c}}&=\frac{1}{2}(G_{c}+G_{d}+m_0^2)\ ,\ \  \frac{\partial W_1}{\partial K_{d}}=\frac{1}{2}G_{c}\nonumber\\
\frac{\partial W_1}{\partial {\hat K}_c}&=\hat{G}_{c}+\hat{G}_{d}+m\hat{m}\ ,\ \ \ \ \frac{\partial W_1}{\partial \hat{K}_{d}}=\hat{G}_{c}\nonumber\\
\frac{\partial W_1}{\partial {\hat{\hat K}}_c}&=\frac{1}{2}(\hat{\hat{G}}_{c}+\hat{\hat{G}}_{d}+\hat{m}^2)\ , \ \ \frac{\partial W_1}{\partial \hat{\hat{K}}_{d}}=\frac{1}{2}\hat{\hat{G}}_{c} \ ,
\end{align}
and the corresponding direct correlation functions are then obtained by inverting the Ornstein-Zernike equations (\ref{EqOZc}) and (\ref{EqOZd}). 

It is rather obvious that this set of equations leads in general to an analytic behavior of the correlation functions as a function of the source $H$. In particular, the function $\hat{\hat{G}}_d$ does not have a singular term proportional to $\delta(0)$ although this is the expected behavior in the original lattice model, as pointed out in the main text. One can easily see  that the condition for a singular behavior to emerge from Eqs. (\ref{EqA9})-(\ref{EqA11}) is that $\hat{\hat{K}}_c=0$: the quantity $Y_{\pm}(H,\hat{H};h_1,h_2)$ then goes to $\pm \infty$ depending on the sign of $H+h_2\pm k$ and this induces a Heaviside step function in Eq. (\ref{EqA9}) and a Dirac delta when differentiating with respect to $H$. 

We now focus on the behavior along the hysteresis loop where it is sufficient to only keep the two Weiss fields $\hat{K}_c$ and $\hat{\hat{K}}_{d}$ (however, the limit  $\hat{H} \rightarrow  -\infty$ which corresponds to the ascending branch will only be taken at the end of the calculation).  The starting point is the simpler partition function 
\begin{align}
\label{EqZrep0}
{\cal Z}_{rep}(\{H^a\}, \hat{H}^a\})= \int \prod_a d s^a d\hat{s}^a \: e^{\frac{1}{2}\hat{\hat{K}}_{d}\sum_{a,b}\hat{s}^a\hat{s}^b} \: e^{\sum_{a}\{s^a\hat{H}^a+ \hat{s}^a[\hat{K}_{c} s^a+H^a -V'(s^a)]\}} 
\end{align}
which also encompasses  the case of the reference system ($J=0$) where $\hat{K}_c=0$ and $\hat{\hat{K}}_{d}=\Delta$. A single auxiliary (random) field $h$ is now required to eliminate the quadratic dependence on $v=\sum_a \hat{s}^a$.
This leads to
\begin{align}
W_1(H,\hat{H})= \int dh \: p(h){\cal W}(H,\hat{H};h)
\end{align}
where $p(h)$ is a Gaussian distribution with zero mean and variance $\hat{\hat{K}}_d$ (which plays the role of a  `renormalized' disorder) and
\begin{align}
\label{Eq_Wmathcal}
{\cal W}(H,\hat{H};h)&=\ln \int ds \: d\hat{s} \: e^{s\hat{H}+\hat{s}[\hat{K}_c s+H+h-V'(s)]}=\ln \int ds \: e^{s\hat{H}} \delta[(k-\hat{K}_c)s-H-h-k\:\mbox{sgn}s]\nonumber\\
&=-\ln(k- \hat{K}_c)+\hat{H}\frac{H+h-k}{k- \hat{K}_c}+[\frac{k\hat{H}}{k- \hat{K}_c}+\ln \big(2\cosh \frac{k\hat{H}}{k- \hat{K}_c}\big)]\theta(H+h+k)\nonumber\\
&+[\frac{k\hat{H}}{k- \hat{K}_c}-\ln \big(2\cosh \frac{k\hat{H}}{k- \hat{K}_c}\big)]\theta(H+h-k) \ .
\end{align}
As a consequence, we find
\begin{align}
W_1(H,\hat{H})&=-\ln(k- \hat{K}_c)+\hat{H}  \frac{H-k}{k- \hat{K}_c}+\frac{2k\hat{H}}{k- \hat{K}_c}{\cal P}(H+k)\nonumber\\
&+\frac{k}{k- \hat{K}_c}[\ln\big(2\cosh \frac{k\hat{H}}{k- \hat{K}_c} \big)-\frac{k\hat{H}}{k- \hat{K}_c}][{\cal P}(H+k)-{\cal P}(H-k)]
\end{align}
where ${\cal P}(x)=\int_{-\infty}^x p(y) dy=\int_{-x}^{\infty} p(y)dy=\frac{1}{2}[1+\mbox{erf} \frac{x}{\sqrt{2\hat{\hat{K}}_d}}]$.
From this, we readily obtain the magnetizations
\begin{align}
m(H,\hat{H})&=\frac{\partial W_1}{\partial \hat{H}} = \frac{H-k}{k- \hat{K}_c}+\frac{2k}{k- \hat{K}_c}{\cal P}(H+k)+\frac{k}{k- \hat{K}_c}[\tanh \frac{k\hat{H}}{k- \hat{K}_c} -1][{\cal P}(H+k)-{\cal P}(H-k)]\nonumber\\
\hat{m}(H,\hat{H})&=\frac{\partial W_1}{\partial H}=\frac{\hat{H}}{k- \hat{K}_c}+\frac{2k \hat{H}}{k-\hat{K}_c}p(H+k)+[\ln \big(2\cosh \frac{k\hat{H}}{k- \hat{K}_c}\big)-\frac{k\hat{H}}{k- \hat{K}_c}][p(H+k)-p(H-k)] \ ,
\end{align}
and the connected (Green's) correlation functions
\begin{align}
\label{EqGc}
G_c(H,\hat{H})=\frac{\partial m}{\partial \hat{H}}=\frac{k^2}{(k- \hat{K}_c)^2}[1-\tanh^2 \frac{k\hat{H}}{k- \hat{K}_c}][ {\cal P}(H+k)- {\cal P}(H-k)]
\end{align}
\begin{align}
\label{EqhGc}
\hat{G}_c(H,\hat{H})=\frac{\partial \hat{m}}{\partial \hat{H}}=\frac{1}{k- \hat{K}_c} +\frac{2k}{k-\hat{K}_c}p(H+k)+\frac{k}{k- \hat{K}_c}[\tanh \frac{k\hat{H}}{k- \hat{K}_c}-1][p(H+k)-p(H-k)]
\end{align}
\begin{align}
\label{EqhhGc}
\hat{\hat{G}}_c(H,\hat{H})=\frac{\partial \hat{m}}{\partial H}=-\frac{2k\hat{H}}{(k-\hat{K}_c)}\frac{H+k}{\hat{\hat{K}}_d}p(H+k)-[\ln (2\cosh \frac{k\hat{H}}{k- \hat{K}_c} -\frac{k\hat{H}}{k- \hat{K}_c}][\frac{H+k}{\hat{\hat{K}}_d} p(H+k)-\frac{H-k}{\hat{\hat{K}}_d} p(H-k)] \ .
\end{align}

In the limit $\hat H \rightarrow -\infty$,  these expressions simplify to
\begin{align}
\label{Eqmag}
m(H)& = \frac{1}{k- \hat{K}_c}[H-k + 2k {\cal P}(H-k)]\nonumber\\
\hat{m}(H,\hat{H})& \sim \frac{\hat{H}}{k- \hat{K}_c} [1+ 2k p(H-k)] \ ,
\end{align}
and
\begin{align}
G_c(H)&=0\nonumber\\
\hat{G}_c(H)&=\frac{1}{k- \hat{K}_c} [1+ 2k p(H-k)]\nonumber\\
\hat{\hat{G}}_c(H,\hat{H})&\sim \frac{2k\hat{H}}{\hat{\hat{K}}_d (k-\hat{K}_c)} (H-k) p(H-k) \ .
\end{align} 

At zeroth-order of the expansion in number of free replica sums, the disconnected functions are obtained from the derivatives of the second cumulant $W_2(H^a,\hat{H}^a;H^b,\hat{H}^b)=\int dh \: p(h) {\cal W}(H^a,\hat{H}^a;h){\cal W}(H^b,\hat{H}^b;h)-W_1[H^a,\hat{H}^a]\:W_1[H^b,\hat{H}^b]$ with respect to the sources. We first consider $G_d(H^a,\hat{H}^a;H^b,\hat{H}^a)=\partial ^2 W_2(H^a,\hat{H}^a;H^b,\hat{H}^b)/\partial \hat{H}^a\partial \hat{H}^b$ and
\begin{align}
G_d(H^a,\hat{H}^a;H^b,\hat{H}^b)+m(H^a,\hat{H}^a)m(H^b,\hat{H}^b)=\int dh \: p(h) \frac{\partial {\cal W}(H^a,\hat{H}^a;h)}{\partial \hat{H}^a}\frac{\partial {\cal W}(H^b,\hat{H}^b;h)}{\partial \hat{H}^b} \ .
\end{align}
The presence of Heaviside step functions in ${\cal W}(H,\hat{H};h)$ makes the result dependent on the sign of $H^a-H^b$. For simplicity, we set at once $\hat{H}^a=\hat{H}^b=\hat{H}$. We then find
\begin{equation}
\begin{split}
G_d(H^a,\hat{H}^a;H^b,\hat{H}^b)&+m(H^a,\hat{H}^a)m(H^b,\hat{H}^b)\\&=\frac{1}{(k-\hat{K}_c)^2}\big[f^{sym}(H^a,\hat{H};H^b,\hat{H})+g(H^a,\hat{H})\theta(H^b-H^a)+g(H^b,\hat{H})\theta(H^a-H^b)\big]
\end{split}
\end{equation}
where $f^{sym}$ is a symmetric function of $H^a$ and $H^b$,
\begin{align}
f^{sym}(H^a,\hat{H};H^b,\hat{H})&=\hat{\hat{K}}_d+(H^a-k)(H^b-k)\nonumber\\
&+k[1+\tanh \frac{k\hat{H}}{k- \hat{K}_c}]\big[(H^b-k){\cal P}(H^a+k)+(H^a-k){\cal P}(H^b+k)+\hat{\hat{K}}_d[p(H^a+k)+p(H^b+k)]\big]\nonumber\\
&+k[1-\tanh \frac{k\hat{H}}{k- \hat{K}_c}]\big[(H^b-k){\cal P}(H^a-k)+(H^a-k){\cal P}(H^b-k)+\hat{\hat{K}}_d[p(H^a-k)+p(H^b-k)]\big]\nonumber\\
&+k^2[1-\tanh^2 \frac{k\hat{H}}{k- \hat{K}_c}][{\cal P}(H^a-k)+{\cal P}(H^b-k)] \ ,
\end{align}
and
\begin{align}
g(H,\hat{H})=k^2[1+\tanh \frac{k\hat{H}}{k- \hat{K}_c}]^2{\cal P}(H+k)+k^2[1-\tanh \frac{k\hat{H}}{k- \hat{K}_c}]^2{\cal P}(H-k) \ .
\end{align}
Therefore, when $H^a,H^b \rightarrow H$, we obtain
\begin{align}
\label{EqDeltacusp}
G_d(H^a,\hat{H};H^b,\hat{H})+m(H^a,\hat{H})m(H^b,\hat{H})&=\frac{1}{(k-\hat{K}_c)^2}\big[f^{sym}(H,\hat{H};H,\hat{H})+g(H,\hat{H})\nonumber\\
&-\frac{1}{2}\vert H_a-H_b\vert \frac{\partial g(H,\hat{H})}{\partial H}+O((H^a-H^b )^2)\big] \ .
 \end{align}
For $\hat{H}^a=\hat{H}^b=\hat{H} \rightarrow -\infty$, $G_d(H;H)$ is then given by
\begin{equation}
\label{Eq_Gd0RS}
G_d(H;H)=\frac{\hat{\hat{K}}_d[1+ 4kp(H-k)]+ 4 k^2 {\cal P}(H-k)[1-{\cal P}(H-k)]}{(k- \hat{K}_c)^2} 
\end{equation} 
whereas the coefficient of $\vert H^a-H^b\vert$  is equal to
\begin{align}
\label{EqcuspRS}
-\frac{2k^2}{(k-\hat{K}_c)^2}p(H-k) \ .
 \end{align}

The 2-replica correlation function $\hat{\hat{G}}_d(H^a,\hat{H}^a;H^b,\hat{H}^a)=\partial ^2 W_2(H^a,\hat{H}^a;H^b,\hat{H}^b)/\partial H^a\partial H^b$ consists of a regular part and a singular part proportional to $\delta(H^a-H^b)$,
\begin{align}
\label{EqhhGd}
\hat{\hat{G}}_d(H^a,\hat{H}^a;H^b,\hat{H}^b)=\hat{\hat{G}}_d^{reg}(H^a,\hat{H}^a;H^b,\hat{H}^b)+\hat{\hat{G}}_d^{sing}(H^a,\hat{H}^a;H^a,\hat{H}^b)\delta(H^a-H^b) \ .
 \end{align}
We find
\begin{align}
\hat{\hat{G}}_d^{reg}(H^a,\hat{H}^a;H^b,\hat{H}^b)+\hat{m}(H^a,\hat{H}^a)\hat{m}(H^b,\hat{H}^b)=&\frac{\hat{H}^a\hat{H}^b}{[k-\hat{K}_c]^2}\Big[1+k[p(H^a+k)+p(H^a-k)+p(H^b+k)+p(H^b-k)]\Big]\nonumber\\
&+\frac{\hat{H}^a}{k-\hat{K}_c}\ln \big(2\cosh \frac{k\hat{H}^b}{k- \hat{K}_c}\big)[p(H^b+k)-p(H^b-k)]\nonumber\\
&+\frac{\hat{H}^b}{k-\hat{K}_c}\ln \big(2\cosh \frac{k\hat{H}^a}{k- \hat{K}_c}\big)[p(H^a+k)-p(H^a-k)]
\end{align}
 and
\begin{align}
\hat{\hat{G}}_d^{sing}(H^a,\hat{H}^a;H^a,\hat{H}^b)=&[\frac{k\hat{H}^a}{k- \hat{K}_c}+\ln \big(2\cosh \frac{k\hat{H}^a}{k- \hat{K}_c}\big)][\frac{k\hat{H}^b}{k- \hat{K}_c}+\ln \big(2\cosh \frac{k\hat{H}^b}{k- \hat{K}_c}\big)]p(H^a+k)\nonumber\\
&[\frac{k\hat{H}^a}{k- \hat{K}_c}-\ln \big(2\cosh \frac{k\hat{H}^a}{k- \hat{K}_c}\big)][\frac{k\hat{H}^b}{k- \hat{K}_c}-\ln \big(2\cosh \frac{k\hat{H}^b}{k- \hat{K}_c}\big)]p(H^a-k)\ .
\end{align}
(There are also singular contributions proportional to  $\delta(H^b-H^a\pm 2k)$ which can be discarded as we are only interested in the vicinity of $H^a=H^b$.) For $H^a,H^b \rightarrow H$ and $\hat{H}^a,\hat{H}^b \rightarrow -\infty$, we then find
\begin{align}
\label{EqGd0reg}
\hat{\hat{G}}_d^{reg}(H,\hat{H}^a;H,\hat{H}^b)+\hat{m}(H,\hat{H}^a)\hat{m}(H,\hat{H}^b)\sim \frac{\hat{H}^a\hat{H}^b}{(k-\hat{K}_c)^2}[1+4kp(H-k)] 
\end{align}
and
\begin{align}
\label{EqGd0sing}
\hat{\hat{G}}_d^{sing}(H,\hat{H}^a;H,\hat{H}^b)\sim \frac{4k^2\hat{H}^a\hat{H}^b}{(k-\hat{K}_c)^2}p(H-k).
\end{align}

Finally, the  2-replica correlation function $\hat{G}_d(H^a,\hat{H}^a;H^b,\hat{H}^a)=\partial ^2 W_2(H^a,\hat{H}^a;H^b,\hat{H}^b)/\partial H^a\partial \hat{H}^b$ is given by
\begin{align}
&\hat{G}_d(H^a,\hat{H}^a;H^b,\hat{H}^b)+\hat{m}(H^a,\hat{H}^a)m(H^b,\hat{H}^b)=\nonumber\\
&\frac{k\: \hat{H}^a}{(k-\hat{K}_c)^2}\Big[\frac{H^b-k}{k}+[1+\tanh \frac{k\hat{H}^b}{k- \hat{K}_c}]{\cal P}(H^b+k)+[1-\tanh \frac{k\hat{H}^b}{k- \hat{K}_c}]{\cal P}(H^b-k)\Big]\nonumber\\
&+\frac{k\:p(H^a+k)}{k- \hat{K}_c}\Big[\frac{k\hat{H}^a}{k-\hat{K}_c}+\ln \big(2\cosh \frac{k\hat{H}^a}{k-\hat{K}_c}\big)\Big]\Big[\frac{H^b-H^a-2k}{k}+[1+\tanh \frac{k\hat{H}^b}{k- \hat{K}_c}]\theta(H^b-H^a)\nonumber\\
&+[1-\tanh \frac{k\hat{H}^b}{k- \hat{K}_c}]\theta(H^b-H^a-2k)\Big]+\frac{k\:p(H^a-k)}{k- \hat{K}_c}\Big[\frac{k\hat{H}^a}{k- \hat{K}_c}-\ln \big(2\cosh \frac{k\hat{H}^a}{k-\hat{K}_c}\big)\Big]\Big[\frac{H^b-H^a}{k} \nonumber\\
&+[1-\tanh \frac{k\hat{H}^b}{k- \hat{K}_c}]\theta(H^b-H^a)+[1+\tanh \frac{k\hat{H}^b}{k- \hat{K}_c}]\theta(H^b-H^a+2k)\Big] \ .
\end{align}
This function has a step-discontinuity at $H^a=H^b=H$ (and also at $H^a=H^b\pm 2k$) and therefore one must fix the value of $\theta(0)$ to lift the ambiguity when the sources are equal. This is done by imposing the exact symmetry $\hat{G}_d(-H,-\hat{H};-H,-\hat{H})=\hat{G}_d(H,\hat{H};H,\hat{H})$ which yields $\theta(0)=1/2$. As a result, one has 
\begin{align}
&\hat{G}_d(H,\hat{H};H,\hat{H})+\hat{m}(H,\hat{H})m(H,\hat{H})=\nonumber\\
&\frac{k\: \hat{H}}{(k-\hat{K}_c)^2}\Big[\frac{H-k}{k}+[1+\tanh \frac{k\hat{H}}{k- \hat{K}_c}]{\cal P}(H+k)+[1-\tanh \frac{k\hat{H}}{k- \hat{K}_c}]{\cal P}(H-k)\Big]\nonumber\\
&+\frac{k\:p(H+k)}{2(k- \hat{K}_c)}[\frac{k\hat{H}}{k-\hat{K}_c}+\ln \big(2\cosh \frac{k\hat{H}}{k-\hat{K}_c}\big)[-3+\tanh \frac{k\hat{H}}{k- \hat{K}_c}]\nonumber\\
&+\frac{k\:p(H-k)}{2(k- \hat{K}_c)}[\frac{k\hat{H}}{k- \hat{K}_c}-\ln \big(2\cosh \frac{k\hat{H}}{k-\hat{K}_c}\big)][3+\tanh \frac{k\hat{H}}{k- \hat{K}_c}] \ ,
\end{align}
so that
\begin{align}
\label{Gdasymp}
\hat{G}_d(H,\hat{H};H,\hat{H})\sim \hat{H}\: \frac{2kp(H-k)}{(k-\hat{K}_c)^2}[2k-H+2k{\cal P}(H-k)] 
\end{align}
for $\hat{H}\rightarrow -\infty$.

Finally, we consider  the two derivatives that are needed in Eq. (\ref{Gd0_cusp_final}) to compute $\partial G_{d}(H^a;H^b)/\partial(H^a-H^b)\big|_{0^+}$ in the single-site effective model in the limit  $\hat{H}^a,\hat{H}^b\rightarrow -\infty$. The first one is simply given by Eq. (\ref{EqcuspRS}),
\begin{align}
\frac{\partial G_{d}}{\partial(H^a-H^b)}\bigg|_{\hat K_c,\hat{\hat K}_d;0^+} = -\frac{2k^2}{[k-\hat{K}_c]^2}p(H-k)\ .
\end{align}

The second one is obtained from  Eq. (\ref{EqGd0reg}),
\begin{equation}
\begin{split}
\frac{\partial G_{d}}{\partial \hat{\hat K}_d(H^a;H^b)}\bigg|_{0^+} &=\lim_{\hat{H}^a,\hat{H}^b\rightarrow -\infty} \frac{\partial^2}{\partial \hat{H}^a\partial \hat{H}^b}[\hat{\hat{G}}_{d}(H^a,\hat H^a;H^b,\hat H^b)+\hat m(H^a,\hat H^a) \hat m(H^b,\hat H^b)]\\&=\frac{1+ 4k p(H-k)}{(k- \hat{K}_c)^2} \ ,
\end{split}
\end{equation}
where we recall that the subscript $0^+$ indicates the limit of equal fields, $H^a=H^b=H$.


\begin{thebibliography}{10}
\bibitem{note1} The fact that there are no metastable states outside the saturation hysteresis loop is a trivial consequence of the so-called `no-passing' rule for ferromagnetic systems\cite{S1993,M1992}. On the other hand, to the best of our knowledge, there is yet no general proof that the complexity in the large-disorder regime is strictly positive {\it everywhere} inside the loop in the large-disorder regime (within the formalism developed in this work, it may very well be that the complexity vanishes at some finite value of the auxiliary field $\hat{H}$, say $\hat{H}^*$ ; this would imply that the number of metastable states does not grow exponentially with the system size for $\hat{H}>\hat{H}^*$  and therefore that the hysteresis loop, obtained in the limit $\hat{H} \rightarrow \pm \infty$, is not the convex hull of the metastable states). However, this seems quite unlikely, as illustrated  by recent  analytical and numerical calculations\cite{DRT2005,PRT2008,RTP2009,RM2009}. On the other hand, in the low-disorder regime, it is conjectured that the region where the complexity is positive is non-convex, and the loss of convexity results in the macroscopic jump in the non-equilibrium magnetization curve (since the response function $m(H)$ must always be convex). Moreover, in this case, it may be that the complexity does not vanish along the reentrant part of the envelope\cite{RTP2009}.
\bibitem{note2} As noted in Ref.\cite{DS1996} (reference 100) $S_{\alpha}$ is not exactly equal to $2N_{\alpha}$ where $N_{\alpha}$ is the number of spins that flip in the avalanche, {\it i.e.} the number of spins that move from the `down' to the `up'  potential well. There is also a contribution to $S_{\alpha}$ that comes from the harmonic response of the other spins through their coupling to the flipping spins.
\bibitem{note3} Note that the Weiss fields $K^{ab}$'s can be considered as sources acting on the composite variables $s^as^b$ (and similarly with hatted quantities) just as $H_a, \hat{H}_a$ are sources acting on the fundamental variables $\hat{s}^a, s^a$. This allows one to generate the (two-particle irreducible) Luttinger-Ward functional for the single-site effective model, which is then related to the local part of the Luttinger-Ward functional of the original model under the condition that the magnetizations $m_a, \hat{m}_a$ and the local part of the Green's functions are the same in the two models.
\end{thebibliography}
 \end{document}